\documentclass[aps,prd,unsortedaddress,superscriptaddress,showpacs,twocolumn,floatfix,nofootinbib]{revtex4-1}

\RequirePackage{lineno}

\usepackage{amsmath}
\usepackage{graphicx}
\usepackage{footnote}
\usepackage{multirow}

\begin{document}

\title{Precision muon decay measurements and improved constraints on the weak interaction}

\affiliation{University of Alberta, Edmonton, Alberta, T6G 2J1, Canada}
\affiliation{University of British Columbia, Vancouver, British Columbia, V6T 1Z1, Canada}
\affiliation{Kurchatov Institute, Moscow, 123182, Russia}
\affiliation{University of Montreal, Montreal, Quebec, H3C 3J7, Canada}
\affiliation{University of Regina, Regina, Saskatchewan, S4S 0A2, Canada}
\affiliation{Texas A\&M University, College Station, Texas 77843, USA}
\affiliation{TRIUMF, Vancouver, British Columbia, V6T 2A3, Canada}
\affiliation{Valparaiso University, Valparaiso, Indiana 46383, USA}

\author{A.~Hillairet}
\email{ant@uvic.ca}
\altaffiliation[Present address: ]{University of Victoria, Victoria, British Columbia.}
\affiliation{TRIUMF, Vancouver, British Columbia, V6T 2A3, Canada}

\author{R.~Bayes}
\altaffiliation[Present address: ]{University of Glasgow, Glasgow, United Kingdom.}
\affiliation{TRIUMF, Vancouver, British Columbia, V6T 2A3, Canada}

\author{J.F.~Bueno}
\affiliation{University of British Columbia, Vancouver, British Columbia, V6T 1Z1, Canada}

\author{Yu.I.~Davydov}
\altaffiliation[Present address: ]{JINR, Dubna, Russia}
\affiliation{TRIUMF, Vancouver, British Columbia, V6T 2A3, Canada}

\author{P.~Depommier}
\affiliation{University of Montreal, Montreal, Quebec, H3C 3J7, Canada}

\author{W.~Faszer}
\affiliation{TRIUMF, Vancouver, British Columbia, V6T 2A3, Canada}

\author{C.A.~Gagliardi}
\affiliation{Texas A\&M University, College Station, Texas 77843, USA}

\author{A.~Gaponenko}
\altaffiliation[Present address: ]{Fermi National Accelerator Laboratory, Ilinois 60510, USA.}
\affiliation{University of Alberta, Edmonton, Alberta, T6G 2J1, Canada}

\author{D.R.~Gill}
\affiliation{TRIUMF, Vancouver, British Columbia, V6T 2A3, Canada}

\author{A.~Grossheim}
\affiliation{TRIUMF, Vancouver, British Columbia, V6T 2A3, Canada}

\author{P.~Gumplinger}
\affiliation{TRIUMF, Vancouver, British Columbia, V6T 2A3, Canada}

\author{M.D.~Hasinoff}
\affiliation{University of British Columbia, Vancouver, British Columbia, V6T 1Z1, Canada}

\author{R.S.~Henderson}
\affiliation{TRIUMF, Vancouver, British Columbia, V6T 2A3, Canada}

\author{J.~Hu}
\altaffiliation[Present address: ]{AECL, Mississauga, Ontario, Canada.}
\affiliation{TRIUMF, Vancouver, British Columbia, V6T 2A3, Canada}

\author{D.D.~Koetke}
\affiliation{Valparaiso University, Valparaiso, Indiana 46383, USA}

\author{R.P.~MacDonald}
\affiliation{University of Alberta, Edmonton, Alberta, T6G 2J1, Canada}

\author{G.M.~Marshall}
\affiliation{TRIUMF, Vancouver, British Columbia, V6T 2A3, Canada}

\author{E.L.~Mathie}
\affiliation{University of Regina, Regina, Saskatchewan, S4S 0A2, Canada}

\author{R.E.~Mischke}
\affiliation{TRIUMF, Vancouver, British Columbia, V6T 2A3, Canada}

\author{K.~Olchanski}
\affiliation{TRIUMF, Vancouver, British Columbia, V6T 2A3, Canada}

\author{A.~Olin}
\altaffiliation[Affiliated with: ]{University of Victoria, Victoria, British Columbia, Canada.}
\affiliation{TRIUMF, Vancouver, British Columbia, V6T 2A3, Canada}

\author{R.~Openshaw}
\affiliation{TRIUMF, Vancouver, British Columbia, V6T 2A3, Canada}

\author{J.-M.~Poutissou}
\affiliation{TRIUMF, Vancouver, British Columbia, V6T 2A3, Canada}

\author{R.~Poutissou}
\affiliation{TRIUMF, Vancouver, British Columbia, V6T 2A3, Canada}

\author{V.~Selivanov}
\affiliation{Kurchatov Institute, Moscow, 123182, Russia}

\author{G.~Sheffer}
\affiliation{TRIUMF, Vancouver, British Columbia, V6T 2A3, Canada}

\author{B.~Shin}
\altaffiliation[Affiliated with: ]{University of Saskatchewan, Saskatoon, Saskatchewan, Canada.}
\affiliation{TRIUMF, Vancouver, British Columbia, V6T 2A3, Canada}

\author{T.D.S.~Stanislaus}
\affiliation{Valparaiso University, Valparaiso, Indiana 46383, USA}

\author{R.~Tacik}
\affiliation{University of Regina, Regina, Saskatchewan, S4S 0A2, Canada}

\author{R.E.~Tribble}
\affiliation{Texas A\&M University, College Station, Texas 77843, USA}

\collaboration{TWIST Collaboration}
\noaffiliation

\date{\today}

\newcommand{\e}[1]{\ensuremath{\times 10^{#1}}}
\newcommand{\pmupixi}{\ensuremath{P_\mu^\pi \xi}}
\newcommand{\pmupi} {\ensuremath{P_\mu^\pi}\,}
\newcommand{\pmuxi} {\ensuremath{\pmupi\xi}}
\newcommand{\cm}    {\,\textrm{cm}}
\newcommand{\mrad}  {\,\textrm{mrad}}
\newcommand{\tes}   {\,\textrm{T}}
\newcommand{\MeV}   {\,\textrm{MeV}}
\newcommand{\MeVc}  {\, \textrm{MeV}/\textit{c}}
\newcommand{\GeVcc} {\,\textrm{GeV}/\textit{c}^2}
\newcommand{\keV}   {\,\textrm{keV}}
\newcommand{\mum}   {\, \mu \textrm{m}}
\newcommand{\mus}   {\, \mu \textrm{s}}
\newcommand{\ns}    {\,\textrm{ns}}
\newcommand{\mr}    {\, \textrm{mrad}}
\newcommand{\abs}[1]{\ensuremath{\left\lvert#1\right\rvert}} 
\newcommand{\ct}    {\ensuremath{\cos\theta}}
\newcommand{\musr}  {$\mu^+$SR}
\newcommand{\pms}   {\ensuremath{\,\textrm{ms}^{-1}}}
\renewcommand{\deg} {\ensuremath{^\circ}}  

\begin{abstract}
  The TWIST Collaboration has completed its measurement of the three
  muon decay parameters $\rho$, $\delta$, and $P_\mu\xi$.
  This paper describes our 
  determination of $\rho$, which governs the shape
  of the overall momentum spectrum, and $\delta$, which controls the
  momentum dependence of the parity-violating decay asymmetry.  The
  results are $\rho=0.749\,77\pm 0.000\,12(\text{stat.})\pm
  0.000\,23(\text{syst.})$ and $\delta = 0.750\,49\pm 0.000\,21(\text{stat.})\pm
  0.000\,27(\text{syst.})$. These are consistent with the value of $3/4$
  given for both parameters in the standard model,
  and each is over a factor of 10 more precise than the measurements
  published prior to TWIST.
  Our final results on $\rho$, $\delta$, and $P_\mu\xi$ have been
  incorporated into a new global analysis of all available muon
  decay data, resulting in improved model-independent
  constraints on the possible weak interactions
  of right-handed particles.
\end{abstract}

\pacs{13.35.Bv, 14.60.Ef, 12.60.Cn}

\maketitle

\section{Introduction}
\label{s:intro}
The TWIST experiment is a high-precision search for evidence of contributions
to the charged-current weak interaction beyond those described by the
standard model (SM) of particle physics. We take advantage of the purely
leptonic nature of the decay of the positive muon into a positron and 
two neutrinos, {$\mu^+ \rightarrow e^+ \nu_e \bar \nu_\mu$}, which
can be described to a good approximation as a four-fermion point interaction
and in the SM is mediated by the $W$ boson.

The most general Lorentz-invariant, local, and lepton-number-conserving description
is given by the matrix element
\begin{linenomath}
\begin{equation}
  \label{eq:mudecay_matrixelem}
  M \sim \sum_{\substack{\gamma=S,V,T\\ \epsilon,\mu=L,R\\(n,m)}}
  g_{\epsilon\mu}^\gamma
  \bigl\langle\bar e_{\epsilon} \bigl\vert\Gamma^\gamma\bigr\vert
  (\nu_e)_n \bigr\rangle
  \bigl\langle(\bar\nu_\mu)_m
  \bigl\vert\Gamma_\gamma\bigr\vert \mu_{\mu}\bigr\rangle,
\end{equation}
\end{linenomath}
where each scalar ($S$), vector ($V$), or tensor ($T$) interaction
between $\mu$-handed muons and $\epsilon$-handed 
positrons has an associated coupling constant $g_{\epsilon\mu}^\gamma$ satisfying certain
normalizations and constraints~\cite{Fetscher:1986}.
Only 19 real and independent coupling constants are needed to describe entirely the interaction
because $g_{RR}^T \equiv 0$ and $g_{LL}^T \equiv 0$, and a common phase is not observable.
In the context of the $V-A$ interaction of the SM, all coupling constants are zero except for $g_{LL}^V=1$.
The coupling constants provide the probability $Q_{\epsilon \mu}$ for a $\mu$-handed muon to decay into
an $\epsilon$-handed positron using
\begin{linenomath}
\begin{equation}
  \label{eq:Qem}
  Q_{\epsilon \mu} = \frac{1}{4} \abs{g_{\epsilon\mu}^{S}}^{2} +
  \abs{g_{\epsilon\mu}^{V}}^{2} +
  3(1-\delta_{\epsilon\mu}) \abs{g_{\epsilon\mu}^{T}}^{2},
\end{equation}
\end{linenomath}
where $\delta_{\epsilon\mu}=1$ for $\epsilon=\mu$ and
$\delta_{\epsilon\mu}=0$ for $\epsilon\neq\mu$.
In particular, a model-independent limit on any muon right-handed
couplings \cite{Fetscher:1986,PDG} is determined from the probability
\begin{linenomath}
\begin{align}
    Q_R^\mu & =  \frac{1}{4} \abs{g^S_{LR}}^2 + \frac{1}{4} \abs{g^S_{RR}}^2
    + \abs{g^V_{LR}}^2 + \abs{g^V_{RR}}^2 + 3 \abs{g^T_{LR}}^2.
 \label{eq:QmuR}    
\end{align}
\end{linenomath}
The differential muon decay spectrum~\cite{Michel50}, using the 
notation of Fetscher and Gerber~\cite{PDG}, can be written as
\begin{linenomath}
\begin{eqnarray}
  \frac{d^2\Gamma}{dx\, d\cos\theta_s}
  & = &
  \frac{m_{\mu}}{2\pi^3} W_{e\mu}^4 G_F^2 \sqrt{x^2-x_0^2}
  \left\{ F_{\mathrm{IS}}(x)\right.\nonumber\\
  && + \left. P_\mu\cos\theta_s F_{\mathrm{AS}}(x) \right\} \label{eq:mudecay_michel}
\end{eqnarray}
\end{linenomath}
where $G_F$ is the Fermi coupling constant, $\theta_s$ is the angle
between the muon spin and the positron momentum, $W_{e\mu}\approx
52.8$~MeV is the kinematic maximum positron energy, $x = E_e/W_{e\mu}$
is the positron's reduced energy, $x_0 = m_e/W_{e\mu}$ is the minimum
possible value of $x$, corresponding to a positron of mass $m_e$ at rest, and
$P_\mu$ is the degree of muon polarization at the time of decay.
$P_\mu$ is typically reduced from $P_\mu^\pi$, which is the helicity of
the muon at the time of its production from a pion decay,
due to depolarization undergone by the muon before it decays.

The isotropic and anisotropic parts of the spectrum
\begin{linenomath}
\begin{eqnarray}
  F_{\mathrm{IS}}(x) & = & x(1-x)
  + \cfrac{2}{9}\,\rho \left( 4x^2 - 3x - x_0^2 \right)\nonumber\\
  && +\,\eta\,x_0 (1-x) + F_{\mathrm{IS}}^{\mathrm{RC}}(x), \label{e:Fisotropic}
\end{eqnarray}
\begin{eqnarray}
  F_{\mathrm{AS}}(x) & = & \cfrac{1}{3} \xi\sqrt{x^2-x_0^2}
  \big[ 1 - x
   + \cfrac{2}{3}\,\delta \big( 4x - 3 \nonumber\\
   && + \big( \sqrt{1-x_0^2} - 1 \big) \big) \big]
   + \xi F_{\mathrm{AS}}^{\mathrm{RC}} (x),\label{e:Fanisotropic}
\end{eqnarray}
\end{linenomath}
are parametrized by four muon decay parameters $\rho$, $\eta$, $\delta$, and $\xi$, which are bilinear combinations 
of the coupling constants $g_{\epsilon\mu}^\gamma$. These four parameters, with the addition
of the radiative corrections $F_{\mathrm{IS}}^{\mathrm{RC}}(x)$ and $F_{\mathrm{AS}}^{\mathrm{RC}}(x)$, are sufficient to describe the shape of
the momentum-angle spectrum of the decay positron. 
We analyze the momentum-angle spectrum rather than the energy-angle
spectrum out of convenience and because for these energies the
difference is insignificant.

The introduction of chiral spin 1 fields to the SM
has been investigated \cite{Chizhov94,Chizhov_Review}.
One consequence is
that nonlocal tensor interactions appear, so that $g^T_{LL}$ and $g^T_{RR}$ are
no longer zero. These new couplings can be measured in particular
through the $\delta$ parameter.

Initial and intermediate measurements of $\rho$ and $\delta$ have
already been published \cite{musser:2005,gaponenko:2005,Robpaper}.
This paper presents a detailed description of
the final measurement of the $\rho$ and $\delta$
decay parameters by the TWIST Collaboration reported in
\cite{mischke:2011}. An identical and simultaneous
analysis of the same data
yielded the final $P_\mu\xi$ parameter determination;
a complete description with an emphasis on the systematic
uncertainties specific to $P_\mu\xi$ was presented in \cite{Jamespmuxi}.
The decay parameter $\eta$ was fixed to the global analysis
value of $\eta=0.0036\pm0.0069$ \cite{gagliardi:2005}
because the sensitivity to this parameter is reduced
due to the multiplying factor $x_0 \approx 10^{-2}$.

\section{Experimental setup}
\subsection{TWIST spectrometer}
A brief description of the experimental setup is given here.
A more detailed description of the apparatus can be found in \cite{TWIST_Apparatus:2005}
with the improvements made for this final analysis described in \cite{Jamespmuxi}.

\begin{figure}[!hbt]
	\includegraphics[width=3.4in]{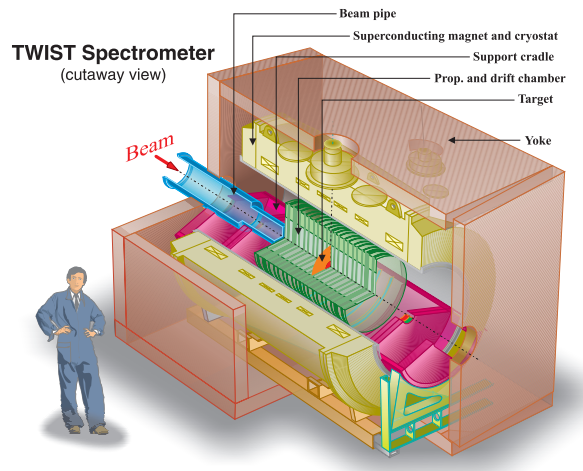}
  \caption{Conceptual drawing of the TWIST spectrometer. It shows the
    superconducting solenoid within the steel yoke, with the drift chambers
    and proportional chambers symmetrically placed upstream and downstream
    of the central stopping target.}
  \label{f:spectrometer-with-man}
\end{figure}

\begin{figure}[!hbt]
	\includegraphics[width=3.4in]{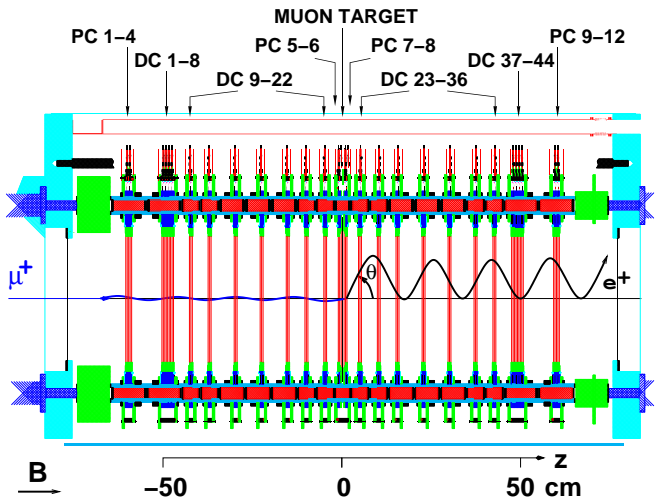}
  \caption{(color online) A cross section of the TWIST detector,
    including an example event of a downstream decay.
    PCs provide timing information and DCs determine the
    position of particles. The angle between the decay positron
	and the $z$ axis, defined along the beam direction,
	is $\theta = \pi - \theta_s$. The muon polarization direction is opposite
	to that of the $z$ axis.}
  \label{f:TWIST_spectrometer}
\end{figure}

An overview of the TWIST apparatus is shown in Fig. \ref{f:spectrometer-with-man};
it was installed on the M13 beamline at TRIUMF, 
Vancouver, Canada. The 500 MeV proton beam from the TRIUMF cyclotron 
hit a carbon target producing pions, some of which stopped and decayed
near the surface of the target to create 29.79 MeV/$c$ muons with 100\% polarization.
The beamline was tuned to transport these highly polarized muons
with a central momentum of 29.6 MeV/$c$ and a momentum bite of 0.7\% FWHM.
The beam also contained several times as many positrons as muons, with the ratio
varying with different tuning conditions.
After passing through the beamline, 
the muons stopped at a rate between 2000 s$^{-1}$ and 5000 s$^{-1}$ 
in a thin target foil located in the center of the 
highly symmetric array of 44 planar drift chambers (DCs) \cite{Davydov} and 12 
planar proportional chambers (PCs) composing the TWIST detector
(Fig. \ref{f:TWIST_spectrometer}).
The DCs and the PCs had an active region of 
32 cm diameter and contained respectively 80 and 160 parallel sense
wires separated by 0.4 cm and 0.2 cm.

The DCs were filled with dimethyl ether gas and were assembled in modules of two or eight chambers 
in which the aluminized Mylar cathode foils were shared by neighboring chambers.
DC 9-22 and 23-36, installed in two-chamber modules,
formed a sparse stack covering most of the tracking
region.
The two eight-chamber DC 1-8 and 37-44 modules instrumented 
the end of the tracking region \cite{Jamespmuxi}. PC 1-4 and 9-12
were installed at the ends of the detector for particle identification purposes.
The PC 5-8 module had the target foil as central cathode foil
and was installed in the center 
of the detector stack to make the entire array symmetric.
The PCs were filled with a mixture of CF$_4$
and isobutane. The array of low mass chambers was installed
in a frame referred to as the cradle, filled with
helium to further reduce the 
amount of material traversed by the muons and positrons.
Two different target foils were used over two run periods to study the effects of 
the target material on the decay parameters measurement:
a (30.9 $\pm$ 0.6) $\mu$m thick silver foil and the 
(71.6 $\pm$ 0.5) $\mu$m thick aluminum
foil used for the intermediate TWIST measurement \cite{Robpaper}.
Both metal targets had purity exceeding 99.999\% and featured
minimal depolarization of the muons after stopping \cite{JamesPRB}.

The detector was installed
in a superconducting solenoid producing a magnetic field of 2 T
that was highly uniform over the tracking region and aligned with
the beam direction. In order to obtain the required field
uniformity and also to reduce fringe fields, it was necessary to
surround the solenoid cryostat with a cube-shaped yoke of
approximately 3 m on a side.
Two NMR probes were installed slightly beyond the radius of the tracking region in
the cradle to monitor constantly the magnetic field strength during data taking.

The magnetic field was mapped using a
rotating arm equipped with Hall probes to measure the longitudinal component
with a precision of 0.1 mT and an NMR probe for the total field.
The Hall probes were separated by about 4.13 cm on the arm.
A full rotation was performed 
every 5.0 cm along $z$ for the central part of the tracking region,
and every 2.5 cm for the edges of the region.
The tracking region was fully mapped for each of the three
field strengths used during data taking, 1.96 T, 2.00 T, and
2.04 T.
A smooth and higher granularity field map, including the relatively small transverse
field components, was calculated using 
the \textsc{Opera}-3D software \cite{opera}, matching the measured magnetic
field map within $\pm$0.2 mT over the drift chamber region.

The beamline vacuum pipe was extended through the fringe field
region as close as possible to the end of the detector
array. Upon exit from the vacuum, muons passed through elements
of a ``beam package,'' including a 20 cm length of gas degrader
filled with an adjustable mixture of He and CO$_2$ gas, a film
strip degrader, and a muon scintillator that triggered the data
acquisition system.  The film
strip degrader consisted of a roll of plastic film containing
holes covered with Mylar degraders of varying thicknesses up to
0.1 cm. It could be rolled from outside the magnet yoke to choose
which degrader was in the muon path. It was used to significantly
degrade the muon beam momentum in order to stop muons well
upstream of the target at the detector center, for special runs
used for positron interaction studies (Sec. \ref{s:US stops}).
The film degrader was set to an empty hole of the film strip for the
normal acquisition of muon decay data. The muons traversed a total
of $\approx$140 mg/cm$^2$ of material, including the beam package and the 
upstream half of the detector, before stopping in the target. The
transverse size of the beam spot was 1.6 cm FWHM.
Because the chambers were operated at atmospheric 
pressure and thus the gas density varied with time,
the ratio between the two gases in the gas
degrader was automatically changed by a feedback loop to set and
maintain the muon stopping distribution in the target.
 
The downstream end of the
detector was equipped with a second beam package during one data
set to test the impact on the data of the asymmetry due to the
presence of the upstream beam package. Two removable time
expansion chambers (TECs) were installed in the beam in the
upstream fringe field region at the beginning and the end of each
data set to characterize the muon beam properties \cite{Hu:2006}.

\subsection{Experimental data}
The data used for the final phase were taken during fall 2006
for the Ag target and in summer 2007 for the Al target 
(see Tables \ref{t:data_sets_Ag} and \ref{t:data_sets_Al};
the numbering of sets is not necessarily sequential).
Monitor information was recorded during all runs for variables such as
spectrometer temperatures, gas pressures and flows, and muon beamline
element settings, and was later evaluated to identify any
instabilities that could signify a low quality of data. 
Approximately 10\% to 30\% of runs in each set were discarded 
prior to the analysis to
guarantee stable run conditions during the period of typically
one week necessary to take a set. The criteria for rejection were
conservative and unbiased; for example, they identified runs with a
problem in the data acquisition system, runs with a noisy
chamber, or runs before the
gas degrader feedback loop was fully locked.

The four nominal sets 74, 75, 84 and 87 were taken with optimal conditions for 
the measurement of decay parameters.
For set 68, the degrader was changed so that the center
of the muon stopping distribution was moved from near the middle
of the target to a point only 1/3 of the way through,
to determine the sensitivity to stopping position variations.
Set 83 was taken with a downstream beam package mirroring the upstream beam package to test 
the impact of the positrons backscattering into the spectrometer and the consistency of the results with 
or without a symmetric apparatus.
Two sets (70 and 71) were taken with different solenoid magnetic field strengths to 
verify that the decay parameters are insensitive to the transverse scale of the helices.

Set 72 was unique in that it was taken with the
TECs in place in the beamline, in order to test the effects
of extra multiple scattering of the muon beam on the parameter
$P_\mu\xi$ through the depolarization of the muons,
and also to monitor the stability of the muon beam position
and angle over an entire week.
The muon beam was steered off the detector axis with an angle $\theta_y 
\approx $30 mrad for set 76 and with a position $x \approx 
$-1 cm and an angle $\theta_x \approx $-10 mrad for 
set 86 to study the depolarization in the fringe field in simulation.
Sets 70, 71, 72,
76 and 86 were discarded from the $P_\mu\xi$ measurement and 
used for systematic uncertainties studies due to their large
depolarization uncertainties \cite{Jamespmuxi}, but
were used for $\rho$ and $\delta$ since these parameters
are insensitive to the muon polarization.

The M13 central momentum was reduced 
to 28.75 MeV/$c$ for set 91 
and to 28.85 MeV/$c$ for sets 92-93 to study the effect of multiple 
scattering of the muons exiting the production target.
The muons were stopped at the entrance of the detector for sets 73 and 80 by changing the momentum 
selection and introducing a film degrader in the beamline. These special sets of data are 
used to validate the simulation (Sec. \ref{sec:MC_validation}).

\begin{table}[!hbt]
		\caption{List of Ag data sets used for the final TWIST measurement.
		The set numbers below are retained for historical reasons;
		missing numbers are not relevant for the analysis.}
		\label{t:data_sets_Ag}
		\begin{tabular*}{1.00\columnwidth}{lccc}
			\hline\hline
			Data   &  Description                  & \multicolumn{2}{c}{Events ($\times 10^6$)} \\
			set    &                               & & \\
			       &                               & Before  & Final  \\
			&                               & cuts   & spectrum \\
			\hline 
			68     & Bragg peak $\tfrac{1}{3}$     &   741  & 32 \\
			& into target & &\\
			70     & Central field at 1.96 T       &   952  & 50 \\
			71     & Central field at 2.04 T       &   879  & 45 \\
			72     & TECs in place, nominal beam   &   926  & 49 \\
			73     & Muons stopped at detector entrance             &   1113 & $\cdots$  \\
			74     & Nominal                       &   580  & 32 \\
			75     & Nominal                       &   834  & 49 \\
			76     & Off-axis beam                 &   685  & 39 \\
			\hline\hline 
		\end{tabular*}
\end{table}

\begin{table}[!hbt]
		\caption{List of Al data sets used for the final TWIST measurement.
		The sets are listed in chronological order except for set 88, which was divided
		into the sets 91, 92 and 93 during the analysis because the running conditions changed.
		The set numbers below are retained for historical reasons;
		missing numbers are not relevant for the analysis.}
		\label{t:data_sets_Al}
		\begin{tabular*}{1.00\columnwidth}{lccc}
			\hline\hline
			Data   &  Description                  & \multicolumn{2}{c}{Events ($\times 10^6$)} \\
			set    &                               & & \\
			       &                               & Before  & Final  \\
			&                               & cuts   & spectrum \\
			\hline
			80     & Muons stopped at              &   363  & $\cdots$ \\
			& detector entrance\\
			83     & Downstream beam package in place                & 943  & 49 \\
			84     & Nominal                         & 1029 & 43 \\
			86     & Off-axis beam                   & 1099 & 58 \\
			87     & Nominal                         & 854  & 45 \\
			91     & Lower momentum I                & 225  & 11 \\
			& $(p=28.75\MeVc)$ \\
			92     & Lower momentum II               & 322  & 15 \\
			& $(p=28.85\MeVc)$ \\
			93     & Lower momentum III              & 503  & 26 \\
			& $(p=28.85\MeVc)$ \\
			\hline\hline 
		\end{tabular*}
\end{table}

\section{Analysis}
The muon decay parameters are extracted from the momentum-angle ($p$-$\theta$) spectrum of the decay positrons 
measured in the TWIST spectrometer. More precisely the difference in shape between the $p$-$\theta$
spectra from the data and from a full simulation of the TWIST apparatus is interpreted 
in terms of a difference in decay parameters.
A blind analysis is performed by using hidden decay parameters for the generation of the 
simulation \cite{AndreiPhD}. These parameters remain hidden until the end of the analysis when all systematic 
uncertainties and corrections have been determined
to minimize the possibility that the results are
affected by human bias.

The simulation is analyzed using the same reconstruction and 
event selection that is applied to the data, and reproduces very closely 
the detector response. Differences between data and simulation arise 
from differences in the muon decay parameters and radiative corrections, 
and additionally from uncertainties in the simulation inputs. The latter 
are the source of most of the systematic uncertainties.

\subsection{Simulation}
\label{s:simulation}
The Monte Carlo simulation of the TWIST experiment uses the \textsc{geant} 3.21 package \cite{Geant321}
to simulate the particle interactions, the detector geometry, and its electronics.
None of the physics processes undergone by the particles such as bremsstrahlung or $\delta$-electron
production are modified or tuned from their definitions in \textsc{geant} 3.21. 
Since our apparatus had very thin scattering layers,
for the energy loss we used the optional simulation
of reduced Landau fluctuations with delta rays.

The simulation includes all the elements necessary to reproduce accurately the muon and positron
trajectories.
The particles are transported in the \textsc{Opera}-3D magnetic field map
using a classical fourth order Runge-Kutta numerical method. 
The description of the wire chambers includes the cathode planes and the wires, as well 
as their positions measured by the alignment calibration (Sec. \ref{sec:DC Calib}).
The discontinuous behavior of the ionization of the wire chamber gas
is simulated with ionization clusters generated randomly along the path
of the charged particles. The ion cluster separation is matched to
the data by comparing the timing of hits close to the wire in data and simulation.
The drift time of each cluster is calculated from DC space-time relations (STRs)
created by a \textsc{Garfield} simulation \cite{Garfield} of the DCs.
The effect of regions of the sense wires becoming temporarily
inefficient due to the presence of
ionization from previous muon hits is also simulated.
The data acquisition digitization is part of the simulation
in order to have output identical in format to
that of the apparatus.

For each data set, a corresponding simulation is generated 
with its input parameters matched to the specific
data taking conditions for that set, as needed.
The fractions of He and CO$_2$ in
the gas degrader are set to time averaged values from the data.
The muon beam profile measured by the TECs is used to
generate the initial muon directions \cite{Jamespmuxi}.
The muon and positron beam rates are matched to the data to
simulate accurately the overlap in time of the hits in the DCs.
Pions and cloud muons\footnote{Cloud muons originate
from pions decaying in flight as they move from the production target
to the M13 beamline. These muons have a low polarization and are therefore
removed during the analysis of the data with a time of flight cut.}
are beam particles that
are not simulated because they can be effectively eliminated from
the experimental data.
The magnetic field strength is matched to
the cradle NMR probe measurements performed during each data set.
Energy loss in some components outside of the tracking region is also
simulated. For example, the upstream beam package
had to be simulated in detail to reproduce
the positrons scattering back into the detector 
and affecting the track reconstruction.
The entire downstream beam package was also
included in the simulation matching set 83.

Individual muons are generated at the location
of the TECs, where the real beam has been well characterized,
with polarization of 100\% in a direction opposite to their
momentum. 
The initial momentum and angle of the decay positron are generated with an independent program in 
order to isolate the hidden parameters of the blind analysis. 
The hidden parameters are chosen randomly within a range of $\pm10^{-2}$ 
from the SM values and remain encrypted during the whole analysis. 
The algorithm uses an accept-reject Monte Carlo technique with the theoretical $p$-$\theta$ 
spectrum including full $\mathcal{O}(\alpha)$ radiative corrections with exact electron mass 
dependence, the leading logarithmic terms of $\mathcal{O}(\alpha^2)$, 
the next-to-leading logarithmic terms of $\mathcal{O}(\alpha^2)$,
leading logarithmic terms of $\mathcal{O}(\alpha^3)$, correction for soft 
pairs and virtual pairs, and an ad-hoc exponentiation \cite{Arbuzov}. 
The $W$ boson's mass and the strong interaction contributions
to the decay through loops are respectively on the order of
$10^{-6}$ and $10^{-7}$~\cite{davydychev01:hadronic_mudecay},
orders of magnitude smaller than our precision goal,
and are therefore ignored for this measurement.

\subsection{Event and track reconstruction}
The reconstruction software is composed of three main algorithms.
It begins by grouping the hits in the spectrometer into different time windows and by identifying
the type of particle (e.g., decay positron,
beam positron, incident muon, secondary electron, etc.) causing
the hits.
Then a pattern recognition algorithm uses the positions of the hit wires
to define helical tracks within each time window,
using spatial information to separate the hits from two particles
completely overlapped in time if necessary.
Electron and positron tracks are finally reconstructed
with high precision using the drift information
in the DCs to extract the momentum and direction of the particles.

Information from a 16 $\mu$s interval is recorded for each event
(from 6 $\mu$s before to 10 $\mu$s after the muon 
trigger) and divided into time windows designed to group
together the signals coming from each particle.
The signals from the PCs define the beginning of the time windows
because their time resolution is $<20$ ns. The time windows 
are by default 1050 ns long to include the longest drift times in the DCs
(50 ns before and 1000 ns after the first PC hit time).
However if two particles 
are separated in time by less than 1000 ns but more than 100 ns, the first time window stops 
at the beginning of the second window.
This type of event is rejected later in the analysis
because signals of the particle in the first window can end up 
in the second window, confusing the track reconstruction.
On the other hand, a time separation of less than 100 ns is not considered long enough for the 
PCs to identify two different particles and only one time window is created.
In this case the signals corresponding to each particle are separated by the pattern recognition
using spatial information. This topology also includes the backscatter of a decay positron from
material outside the tracking region creating two independent tracks overlapping in time,
as well as delta rays emitted in the tracking region.

The particle identification algorithm uses the pulse widths in the PCs,
roughly proportional to the energy deposited, to separate muons from positrons since the two 
particles deposit different amounts of energy.
Beam positrons are identified using the fact that they traverse the entire detector while the decay positrons originate 
from the target foil region in the middle of the chamber stack.
The events are classified according to the particle content and the length of the time windows.

The track reconstruction algorithm is performed on the signals in each time window. 
The first part of this algorithm is a pattern recognition,
which combines hits on adjacent wires and associates
signals together to form a coarse estimate 
of the helical track.
The drift times are ignored at this stage and for this reason
the Chebyshev norm is used as a fit optimizer \cite{F_James}.
This pattern recognition identifies and separates the tracks
from the different particles contained in a time window, including $\delta$-ray 
electrons.
A particle undergoing a large enough scattering or energy loss
due to the emission of a bremsstrahlung photon or a $\delta$-ray 
electron is reconstructed as two individual tracks by the algorithm. 

The next stage of the track reconstruction uses a $\chi^2$ minimization
to refine the helical trajectory identified by the 
pattern recognition.
This helix fitter minimizes the residuals at each DC plane as well as kink angles in the center of 
each DC module, and includes as a fit parameter the decay time of the muon. The time of flight 
of the decay positron to each DC plane is included in this calculation.
The kink approach is well adapted 
to the TWIST spectrometer since the scattering masses are discrete \cite{Lutz}.
The kink angles are weighted in the $\chi^2$ minimization 
by the inverse of the width of the Gaussian approximation calculated using
the formula for multiple scattering through small angles \cite{PDG Thru Matter}. 
For this analysis the space-time relationships used to convert
the drift times into drift distances were measured using decay 
positron tracks (see Sec. \ref{sec:DC Calib}).
The trajectories between the DCs are calculated using the \textsc{Opera}-3D magnetic field map to account for the inhomogeneities of 
the solenoid magnetic field. The algorithm uses an arc step approximation with variable size steps to integrate the magnetic 
field features.
The energy lost by the positron through ionization is taken into account in the fitting procedure using
\begin{linenomath}
\begin{equation}
	\Delta E = \frac{1}{\cos \theta} \sum_i l_i \epsilon_i^{ion}
\end{equation}
\end{linenomath}
with $\Delta E$ the average energy loss of a track segment, $l_i$ the thickness
of the material $i$, and $\epsilon_i^{ion}$ the ionization energy lost per unit of thickness
in the material $i$ calculated from the mean energy loss formulas \cite{PDG Thru Matter}.
The track reconstruction has an inefficiency
of a few $10^{-4}$, and an angle-dependent resolution at the end point (52.8 MeV/$c$),
which is 58 keV/$c$ when extrapolated to $\sin\theta = 1$.
From simulation, the absolute accuracy
of the reconstructed momentum is better than $1 \times 10^{-4}$.

\subsection{Event selection}

It is desirable to select classes of events that are very simple and therefore well simulated
to reduce discrepancies between data and simulation.
Our main selection is to find one muon and one decay positron separated by more than 1 $\mu$s.
Events also containing a beam positron are kept only if the beam particle
is separated from the incident muon and decay positron
by more than 1 $\mu$s or less than 100 ns.
A track from a decay positron backscattering at the upstream end of the detector and
a beam positron track are indiscernible by the particle identification.
The backscattering
depends strongly on the decay positron momentum and angle.
Thus events with a backscattered
positron and events with overlap of decay and beam positrons within
100 ns are included in the analysis.
These choices reduce the sensitivity of
the analysis to the accuracy in the simulation of these processes. 

The highly polarized surface muons
are selected using time of flight of the particles
in the M13 beamline \cite{Jamespmuxi}.
A highly polarized muon beam is crucial for the measurement of $P_\mu\xi$,
but also increases the sensitivity to the $\delta$ parameter.
The muons stopping in the target foil are selected by the next series of cuts.
The first PC downstream and adjacent to the target acts as a veto for muons stopping too far downstream. 
The pulse widths in the two PCs just upstream of the target are used to eliminate muons that stopped 
in the gas or the wires of those chambers \cite{Jamespmuxi}. Also the muon position on the target measured by 
the two PCs upstream is used to reject muons stopping more than 2.5 cm away from the central axis 
of the detector. Decay positrons from these rejected muons might not
be contained within the tracking region.

The purpose of the following selections is to identify which track corresponds to the decay positron. 
Tracks that failed the second stage of the track reconstruction and
tracks corresponding to negatively charged particles are rejected. 
The event classification determined on which side of the target the decay positron was emitted based on the
side containing most of the hits.
Tracks located on the opposite side are discarded. 
The next selection tries to match together tracks
to check whether they originated from the same particle.
In particular, the algorithm tries to match tracks from opposite 
sides of the target (using previously discarded tracks)
to identify beam positron tracks and remove them from the analysis. 
In this case the criteria for a match are a time separation of less than 60 ns for the 
track times and a closest distance of approach of the two extrapolations of the tracks of less than 
0.5 cm. The matching can also identify trajectories split in two tracks (both located on one side of the target)
due to a large scattering in a DC.
In this case the closest distance of approach is only required to be 2 cm. The position at 
the target of the muon as measured by the target PCs is compared to the extrapolation of the positron 
track back to the target to determine the vertex distance. An angle-dependent cut is applied to this vertex 
distance. If more than one track candidate was selected,
two more selections determine a single track corresponding to the decay 
positron.
The tracks that are farthest from the target plane are discarded.
If multiple tracks are equally close to target,
the selected track candidate is the one with the shortest muon-positron vertex distance.
Finally only the decays happening between 1050 ns and 9000 ns are selected. Earlier 
tracks might overlap with DC signals from the muon.
DC signals from later tracks may occur after the end of the event recording.

It is important to recall that exactly the same algorithms
are applied to data and simulation,
reducing the dependence of our muon decay results on the precision of the algorithms.
The evaluation of the systematic uncertainties from the detector
response is accomplished using event selection criteria
identical to that of the analysis, and therefore integrates
the effect of the cuts in the uncertainties (Sec. \ref{sec:SystAndCorr}).

\subsection{Muon decay parameter fit}
The $p$-$\theta$ spectra obtained for data and simulation are now compared to
perform a momentum calibration and to extract 
their difference in terms of decay parameters.
The data and simulation spectra have very different muon decay parameters,
compared to the precision of the measurement, because of the hidden parameters in simulation.
This difference is typically a few parts in $10^{-3}$ and
it biases the edge fit of the momentum calibration 
performed at the kinematic end-point of the two spectra.
The shape of the spectrum near the end-point is sensitive to this difference,
so it is necessary to include 
in the simulation the derivatives weighted according to the results from a prior decay parameter fit.
For this reason the two fitting procedures are applied iteratively, 
starting with the decay parameter fit.
Only one iteration of the momentum calibration is needed to reach convergence.

The muon decay parameter fit procedure exploits the linearity in the decay parameters $\eta$, $\rho$ and the products $P_\mu\xi$ 
and $P_\mu\xi\delta$ [Eq. \eqref{eq:mudecay_michel}].
The difference between the data spectrum ($S_D$) and the Monte Carlo simulation spectrum ($S_{MC}$) can 
be expressed in terms of derivative spectra of the decay parameters \cite{AndreiPhD}. Schematically:
\begin{linenomath}
\begin{eqnarray}
\label{eq:DiffDataMCSpectra}
	S_D = S_{MC} & + & \frac{\partial S}{\partial \rho} \Delta\rho \nonumber\\
				 & + & \frac{\partial S}{\partial P_\mu\xi} \Delta(P_\mu\xi) + \frac{\partial S}{\partial P_\mu\xi\delta} \Delta(P_\mu\xi\delta),
\end{eqnarray}
\end{linenomath}
where the $\Delta\alpha, (\alpha = \rho, P_\mu\xi, P_\mu\xi\delta)$ are the free parameters of the fit.
The effect of the detector response on the $p$-$\theta$ derivative spectra is simulated using the same code as is 
used for the muon decay spectrum. However, unlike the decay spectrum, the derivatives are not positive definite, and 
additional sign information must be passed to the fitting software.
The radiative corrections are already taken into account in the simulation.

\begin{figure}[tbh]
	\centering
	\includegraphics[width=\columnwidth]{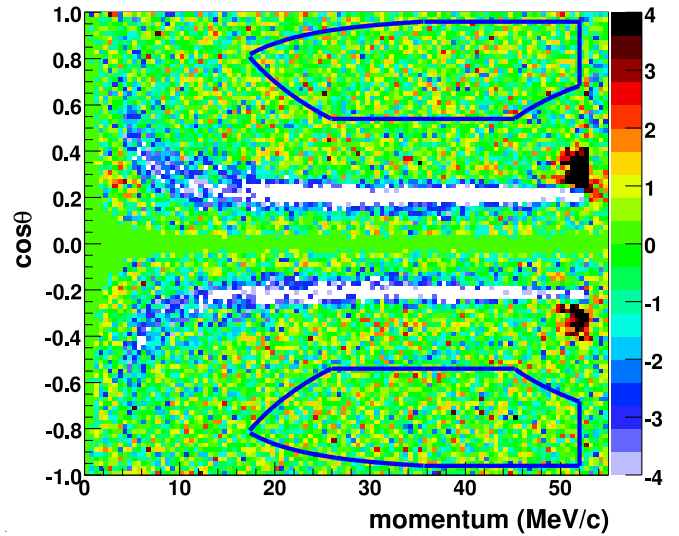}
	\caption{(color online) Residuals normalized by the statistical
	uncertainty from the muon decay parameters fit between simulation and data.
	Only bins with their center contained in the fiducial
	regions are used in the fitting procedure.}
	\label{f:Residuals}
\end{figure}

Fiducial regions in the $p$-$\theta$ spectrum are defined to reduce bias
while maximizing resolution and sensitivities to the decay parameters.
Only the bins whose center is contained in the fiducial regions are used in the decay parameters fit.
The maximum momentum cut ($p^{\textrm{max}} = 52.0\MeVc$) 
avoided the region of the spectrum that 
was used in a momentum calibration procedure
(described below).
The longitudinal momentum cut ($|p_z^{\textrm{min}}| = 14.0\MeVc$)
avoided the region where the helix wavelength was difficult
to determine.
The requirement $|\cos\theta| < 0.96$ removed
small angle tracks where the wavelength
was poorly resolved, and $|\cos\theta| > 0.54$
eliminated large angle tracks with less reliable
reconstruction due to multiple Coulomb scattering as
the path length through the chambers became too large.
The maximum transverse
momentum cut ($p_t^{\textrm{max}} = 38.0\MeVc$) retained only the positrons
within the instrumented regions of the detector.
The minimum transverse momentum cut
($p_t^{\textrm{min}} = 10.0\MeVc$)
removed tracks where the helix radius became
comparable to the wire spacing.
The upstream and downstream fiducial regions are symmetric about $\cos\theta=0$
(Fig. \ref{f:Residuals}).
We studied the stability of the 
decay parameters with respect to the definition of the regions by
varying by a few percent all the fiducial boundaries.
These boundaries were slightly modified for this
analysis compared to the ones used for the intermediate measurement \cite{Robpaper}.

A $\chi^2$ minimization, using \textsc{MINUIT} \cite{MINUIT}, of Eq. \eqref{eq:DiffDataMCSpectra}
to the data is used to 
determine the muon decay parameter differences.
The correlations between the parameters as returned by the fitting algorithm
are 0.19 for $\rho$-$\delta$, 0.21 for $\rho$-$\xi$ and $-0.72$ for $\delta$-$\xi$.
The parameter $\eta$ is not part of the fit in this analysis because it is strongly
correlated to the parameters $\rho$ and $P_\mu\xi$  (Sec. \ref{s:intro}) and
the fiducial regions exclude
the low momentum part of the spectrum, which is the most sensitive to this parameter.
The final determination of $\delta$ from $\Delta(P_\mu\xi\delta)$ is only
possible using the hidden parameters $\delta_h$ and $\xi_h$ in the formula:
\begin{linenomath}
\begin{equation}
	\delta = \frac{\delta_h\xi_h + \Delta(P_\mu\xi\delta)}{\xi_h + \Delta(P_\mu\xi)}.
\end{equation}
\end{linenomath}
However, before unblinding it is sufficient to use the SM values 
to estimate $\Delta \delta$.
The final value of $\delta$ is recalculated after the hidden parameters have been revealed.

\subsection{Momentum calibration}
\label{sec:ECal}
\begin{figure}[tb]
	\centering
	\includegraphics[width=\columnwidth]{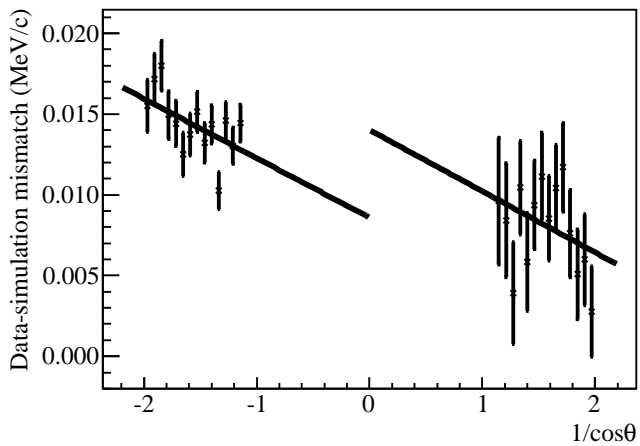}
	\caption{Measurement of the angle-dependent momentum mismatch at the decay
	positron kinematic end point for set 84,
	taken with the Al target under nominal conditions.}
	\label{f:ECal_ShiftVsInvCosTheta}
\end{figure}

The momentum calibration exploits the kinematic end-point of the decay positron momentum
at 52.83 MeV/$c$ 
to measure the mismatch between the data and simulation detector responses. 
Because the planar geometry of the TWIST detector, the momentum loss
of the positrons exiting the target will have a $1/\!\cos\theta$ dependence.
Histograms of the edge region 
with 10 keV/$c$ momentum binning and bins in
$1/\!\cos\theta$ of width 0.0636 in the range
$0.5<|\cos\theta|<0.9$ ($1.11<|1/\!\cos\theta|<2.00$) are produced.
For each $1/\!\cos\theta$ slice the simulated edge histogram is shifted in
10 keV/$c$ steps with respect to the data histogram.
At each step a $\chi^2$ statistic is calculated using the difference in bin contents between the spectra.
The resulting $\chi^2$ distribution is fitted with a second-order polynomial
to determine the momentum shift required to minimize the $\chi^2$.
The momentum mismatch between data and simulation versus
$1/\cos \theta$ (see Fig. \ref{f:ECal_ShiftVsInvCosTheta})
is fitted independently upstream and downstream
with straight lines,
\begin{linenomath}
\begin{equation}
	\Delta p = a_i/\abs{\cos(\theta)} - b_i;\ i=(\mbox{up}, \mbox{dn}).
\end{equation}
\end{linenomath}
A new data $p$-$\theta$\ spectrum is produced by applying the momentum calibration
for each set on an event-by-event basis, and the statistical uncertainties and correlations
of the calibration parameters are propagated to the muon decay parameter error budget.
Table \ref{tab:ECalParam} shows the mean values of the momentum calibration parameters.
\begin{table}[tb]
		\centering
		\caption{Mean values of the momentum calibration parameters
		with statistical uncertainties.}
		\label{tab:ECalParam}
		\begin{tabular*}{1.0\columnwidth}{@{\extracolsep{\fill}}lcccc}
				\hline \hline
				Target	& $a_{up}$ & $b_{up}$ & $a_{dn}$ & $b_{dn}$ \\
                                    & keV/$c$  & keV/$c$  & keV/$c$  & keV/$c$  \\
				\hline
				Ag  	& 1.8 $\pm$ 0.5 &-10.0 $\pm$ 0.8 & -3.1 $\pm$ 1.3 & -1.7 $\pm$ 2.0 \\
				Al  	& 4.8 $\pm$ 0.6 &-6.9  $\pm$ 0.9 & -0.2 $\pm$ 1.4 & -11.0$\pm$ 2.3 \\
				\hline \hline
		\end{tabular*}
\end{table}

The model used for the propagation of the momentum mismatch
to the entire spectrum depends on the source or sources of 
the mismatch, which could not be uniquely identified.
For this reason the final muon decay parameter results are
the average of the 
analyses calibrated using a shift that was either constant
or scaled with momentum.
Systematic uncertainties associated with the momentum
calibration are discussed in Sec. \ref{sec:SystEnergyCalib}.

\subsection{Drift chamber calibration}
\label{sec:DC Calib}
Improvements to the DC calibration procedures have been crucial
to reach our final precision for the decay parameters. 
First of all the wire time offsets, which correct for
the different propagation times of the signals from different 
sense wires, were measured directly from the decay positrons in the physics data.
Previously the wire time offsets were 
determined from special pion data taken only at the
beginning and the end of run periods, leading to 
a dominant systematic uncertainty from the time dependence of these offsets.
For this measurement, a downstream scintillator
was used in addition to the existing upstream scintillator.
Both scintillators 
recorded the arrival time of
the decay positron as a reference.
The upstream scintillator is an annular shaped positron scintillator
installed around the main muon trigger scintillator.
The downstream scintillator on the other hand is installed
outside of the steel yoke and covers most of the yoke downstream opening.

The wire time offsets 
were extracted from the decay positron signals after a time of flight correction.
The algorithm for fitting these time 
distributions, which are broadened by the drift times of the electrons in the DCs,
was significantly improved.
The mismatch of the offsets between data and simulation was estimated
to be less than 0.5 ns channel-by-channel based on 
the difference in the fit parameter describing the steepness of the DC signal rising edge.

The relative misalignment of the DCs was measured and
corrected in the analysis to improve the reconstruction resolution. 
A special set of data was taken with 120 MeV/$c$ pions and with no magnetic field.
The straight tracks 
produced by the pions traveling through the entire chamber stack were reconstructed.
At each wire chamber, the residuals were 
used in an iterative process to determine the misalignment
in translation in the direction measured by the chamber (perpendicular 
to the wires) and in rotation around the detector axis with a precision respectively of 10 $\mu$m and 0.03 mrad. 
The target PCs misalignment was also corrected
due to their importance to measure the muon-positron vertex distance. 
The misalignment between the spectrometer and the magnetic
field axis was measured on muon decay data using a special 
helix fitting algorithm allowing for a rotated helix axis.
This measurement was performed 3 times, each time that the 
spectrometer was removed from inside the coils.
The three misalignments showed remarkable
reproducibility, being consistent within the 0.03 mrad uncertainty, with an 
average value of 0.31 mrad in $x$ and 1.15 mrad in $y$.

The previous TWIST analyses used STRs extracted from \textsc{Garfield} simulations.
This analysis measured effective STRs independently for both simulated
and real data from the time 
residuals of the helix fitter on decay positron tracks \cite{AlexNIM}.
An iterative procedure modified the STRs to 
reduce the time residuals in subcells of the drift cell surrounding the sense wire.
All the drift cells are averaged for each plane.
The main advantages of the new procedure are to correct for a bias from the 
helix fitter, which systematically defines the closest distance of
approach of the track to the wire to be less 
than the actual ion cluster distance to the wire,
and to allow data and simulation
to be treated in a more equivalent way in the analysis.
Furthermore the STRs were measured for each plane in data 
to take into account imperfections in the DCs construction
such as the cathode foil position relative to the wires. 
On the other hand, one set of STRs, measured from the simulation, was applied to all the DCs
in the simulation analysis since in that case
the geometry is identical for all chambers.

The position resolution used during the helix fitting to weight the residuals was
changed from a constant 100 $\mu$m to an \textit{ad hoc} expression determined
by optimizing the momentum bias and resolution in the simulation,
\begin{linenomath}
	\begin{equation}
			\sigma(x) = \left\{120 + 5 \left[ \sinh(100x^2) \right]\right\} \mu\mathrm{m},
			\label{e:vcr}
	\end{equation}
\end{linenomath}
where $x$ is the distance between the wire and the ionization in cm. 
Equation \eqref{e:vcr} assigns a larger uncertainty to hits that are far from the wire, which are affected more by 
diffusion. For $x<0.1\cm$ there is little sensitivity to the position resolution function since a left-right
ambiguity\footnote{Only the drift time and therefore the distance to the wire are known.
In those conditions, the left-right ambiguity corresponds to the difficulty for the reconstruction algorithm
to determine on which side of the sense wire the track of the particle occurs.} dominates.
The improved resolution dependence modified the weights used for the track 
fitting and resulted in a difference between
data and simulation momentum resolutions of $<$ 2 keV/$c$
at the kinematic end-point.

\section{Validations}
\label{sec:MC_validation}

Many low-level histograms, such as distributions of chamber hits and track lengths,
were examined to ensure that the simulation accurately reproduced the data. Very
little tuning of the simulation was required. As mentioned above, none of the
physics processes were tuned from their GEANT defaults. Because the systematic
uncertainty for positron interactions was a leading term for our intermediate results,
this section includes a detailed description of the results of special data taken 
to test the ability of the simulation to reproduce
positron interactions in the detector.  These data also allow a precision
test of reconstruction inefficiencies in data and simulation.  A third subsection 
describes time spectrum fits, which tested the purity of the events in the fiducial
region as positrons from muon decay.

\subsection{Positron interactions}
\label{s:US stops}
A special data set where the muons are stopped far upstream in the muon 
counter and upstream PCs before reaching the DCs is used to validate the relevant positron 
interactions in the simulation, independent of the muon decay parameters. 
In this configuration, a positron from a muon decay traverses the entire 
detector and provides two track segments, one on each side of the target. 
The comparison is restricted to single upstream and downstream 
tracks with hits on at least 16 DC planes and on an outer PC plane. 
The positron track is also required to pass within 4 cm of the target center, 
which limits the $p$-$\theta$\ phase space over which this comparison can be 
made. The fitted tracks return the position and momentum 
at the drift chamber nearest to the stopping target.  
\begin{figure}[tb]
	\centering
	\includegraphics[width=\columnwidth]{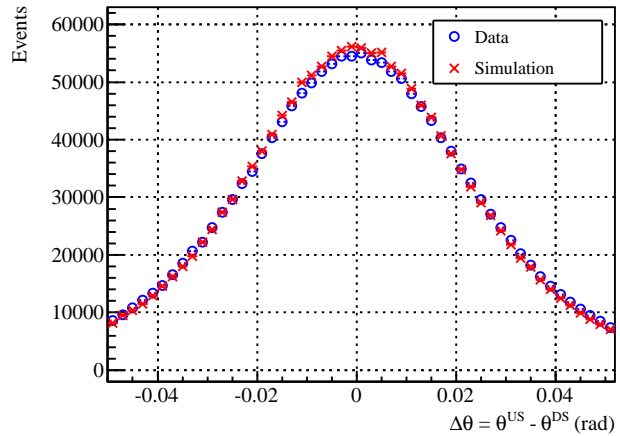}
	\caption{(color online) Integrated $\Delta \theta$ distributions for the silver 
                 target module in data and simulation.}
	\label{f:Ag_DeltaCosTheta}
\end{figure}
The difference in angle between the two reconstructed tracks provides a
test of the ability of the simulation to reproduce 
multiple scattering through the target module. The
distribution of the change in angle is presented for the silver target module
in Fig. \ref{f:Ag_DeltaCosTheta}. The central width and most probable value 
(MPV) of this distribution are obtained
from a fit to a Gaussian function. To minimize the effects of
non-Gaussian tails to this value, the fit region is restricted to $\pm1\sigma$
about its central value.
The agreement in both width and MPV is shown in Table  \ref{t:diff_dist_prop}.

\begin{table*}[!tbh]
  \caption{Properties of integrated  momentum loss ($\Delta p |\cos\theta|$) 
           and scattering ($\Delta \theta$) distributions. The peak and width 
           of the distributions were determined using a truncated Gaussian to 
           remain independent of the long, asymmetric tails. Only
		   statistical uncertainties are quoted.}
  \label{t:diff_dist_prop}
  \begin{tabular*}{2.05\columnwidth}{@{\extracolsep{\fill}}lcccccccc}
  \hline \hline
   & \multicolumn{4}{c}{Silver} & \multicolumn{4}{c}{Aluminium} \\
  & \multicolumn{2}{c}{$\Delta p |\cos\theta|$} & \multicolumn{2}{c}{$\Delta \theta$} & \multicolumn{2}{c}{$\Delta p |\cos\theta|$} & \multicolumn{2}{c}{$\Delta \theta$}\\
  &  Peak keV/$c$ & Width keV/$c$ & Peak mrad & Width mrad & Peak keV/$c$ & Width keV/$c$ & Peak mrad & Width mrad \\
  \hline
  Data  & 40.37 $\pm$ 0.46   & 55.46 $\pm$ 0.20  & -0.00 $\pm$ 0.14  & 21.09 $\pm$ 0.08 & 32.25 $\pm$ 0.42 & 53.28 $\pm$ 0.26 &  0.13 $\pm$ 0.15 & 11.43 $\pm$ 0.06\\
  Simulation  & 43.36 $\pm$ 0.43 & 54.84 $\pm$ 0.26 & -0.20 $\pm$ 0.11 & 20.65 $\pm$ 0.10 & 32.98 $\pm$ 0.57 & 52.21 $\pm$ 0.25 & -0.09  $\pm$ 0.12 & 11.30 $\pm$ 0.05 \\
  \hline \hline
  \end{tabular*}
\end{table*}

\begin{figure}[tbh]
	\centering
	\includegraphics[width=\columnwidth]{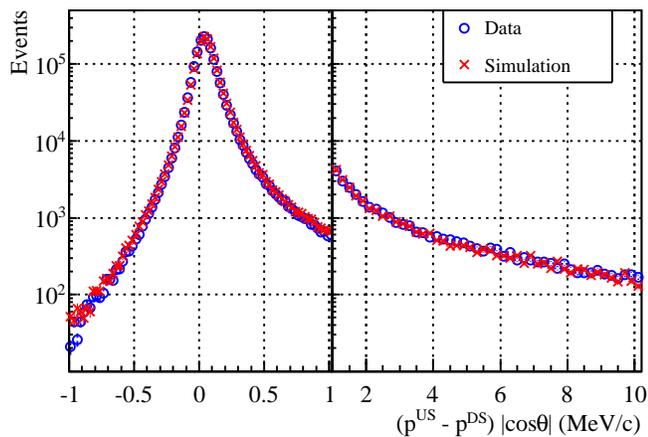}
	\caption{(color online) Integrated momentum loss
	$(\Delta p)|\cos\theta|$ distributions for the silver
	target module in data and simulation. The right panel 
	shows the distribution of high momentum loss events, due primarily to bremsstrahlung.
	The discontinuity between the two panels is because of the change in bin size.}
	\label{f:Ag_Deltap}
\end{figure}

The second measurement comes from the change in momentum, which tests the 
validity of the simulation's positron momentum loss. The measured momentum difference 
shows a $1/\!\cos\theta$ dependence due to the planar geometry of the detector. 
The momentum loss is therefore studied  using the quantity 
$\Delta p |\cos\theta|$ as shown for the silver target module in 
Fig. \ref{f:Ag_Deltap}.
Again a truncated Gaussian fit is used to determine the
MPV and width of the central peak, which measures soft
momentum loss processes. For Al there is agreement at the
1 keV/$c$ level. For Ag there is a 3 keV/$c$ difference in
the MPV momentum loss, which is within the uncertainty of 3.5 keV/$c$ for the
simulated positron momentum loss \cite{Berger}.
The high momentum loss tail extending 10 MeV/$c$ above the peak and 3 orders of magnitude 
below the peak height is due primarily to bremsstrahlung processes. The 
integrated counts in this tail validate the bremsstrahlung rate at the 1\% 
level, in agreement with a separate evaluation based on broken 
tracks (Sec. \ref{sec:SystBremDelta}). 

\subsection{Reconstruction Inefficiencies}
\label{s:US-DS inef}
The analysis of the far upstream stops data also determines the
probability of not finding a track in one-half of the detector when it
is successfully reconstructed in the other half. This inefficiency
includes the possibility that the tracks physically scatter into or
out of the fiducial region but it is dominated by the possibility of
not reconstructing an existing track. The double difference between
upstream and downstream halves of the detector and between data and
simulation would affect the muon decay parameters,
$P_{\mu}\xi$ in particular.

A weighted average of the track inefficiency was compiled 
from the events that fall within the fiducial region for both data and the 
simulation for each target module (Table \ref{tab:ineff}). The weighting was
defined using the Bayesian interval
for the ratio of the failed tracks over the total tracks for a given bin.
Beam positron 
tracks are localized at the $\cos\theta = 0.94$ fiducial boundary and 
are therefore rather sensitive to inscattering and outscattering; for this 
reason they were removed from the calculation. Table \ref{tab:ineff} shows 
a clear difference in the upstream and downstream inefficiencies due to 
positron interactions in the target, but is reproduced by the simulation 
at the  $0.5\times 10^{-4}$ level. Positron interactions in the target module
or first downstream chambers, including annihilation-in-flight,
large angle scattering, or production of secondaries that confound the
reconstruction, will produce such a difference in our inefficiency measurements. 
\begin{table}[tbh]
		\centering
		\caption{Weighted average track inefficiencies upstream
		(US) and downstream (DS) of the target within
		the fiducial region used for the decay parameter fit.
		Only statistical uncertainties are quoted.}
		\label{tab:ineff}
		\begin{tabular*}{0.95\columnwidth}{@{\extracolsep{\fill}}lcccc}
				\hline \hline
				Target				& Detector	& \multicolumn{3}{c}{Inefficiency ($\times 10^{-4}$)} \\
									& half	& Simulation		& Data				& Difference \\
				\hline
				\multirow{2}{*}{Al}	& US	& 3.96 $\pm$ 0.16	& 3.74 $\pm$ 0.16	&  0.36 $\pm$ 0.23 \\
									& DS	& 5.71 $\pm$ 0.18	& 6.15 $\pm$ 0.19	& -0.30 $\pm$ 0.28 \\
				\multirow{2}{*}{Ag}	& US	& 4.54 $\pm$ 0.16	& 3.74 $\pm$ 0.11	& -0.30 $\pm$ 0.20 \\
									& DS	& 7.13 $\pm$ 0.18	& 7.47 $\pm$ 0.15	& -0.58 $\pm$ 0.25 \\
				\hline \hline
		\end{tabular*}
\end{table}

\subsection{Time Spectrum Fits}
To check the consistency of data and simulation, of time
calibration, and the absence of time-independent backgrounds
in the data, fits of the selected events to the time
dependence were performed for a typical data set and also for a
simulation set.
The fits included an overall normalization, the degree of
initial muon polarization, and also
a small time-dependent relaxation of the asymmetry \cite{Jamespmuxi}.
The fit
range was from 2 $\mu$s to 9 $\mu$s following muon arrival
to avoid a small decay time distribution bias below 1 $\mu s$
from the algorithm that rejected beam positron pileup.
Assuming zero uniform background and the accepted value of the muon
lifetime,
acceptable fit qualities were obtained for events in the decay
parameter fit region.
The confidence levels are 75\% for set 84 and 6\% for the
corresponding simulation, using only statistical uncertainties.
These results confirm that the tracks selected by the fiducial
region in data are consistent with a pure sample of positrons
from muon decays. However, no systematic evaluation of the lifetime
measurement was attempted, as it was beyond the scope of our
physics goals and not intrinsically relevant to the measurement
of decay parameters.

\section{Blind analysis uncertainties and corrections}
\label{sec:SystAndCorr}

Most of the systematic uncertainties originate from a mismatch in the apparatus or in physics processes between the simulation 
and the experiment.
These uncertainties are evaluated by purposely exaggerating
the mismatch in a simulation and measuring the change in decay parameters
between this modified simulation and a nominal simulation.
The difference is the \textit{sensitivity} of the decay parameter to this 
mismatch.
The exaggeration produces statistically well determined 
sensitivities.
A factor corresponding to the ratio between the exaggerated
mismatch and the estimation of the real mismatch is used 
to rescale the sensitivity to obtain the systematic uncertainty.
The sensitivity to a component of the analysis can be obtained by
comparing via a decay parameter fit the
spectra from a standard analysis and from an analysis with that component
exaggerated, using the same data for both analyses. This approach, when possible,
reduces the statistical uncertainties from the sensitivity evaluation.
It relies on the assumption of linearity of the systematic uncertainties
evaluated, which was verified to be valid for large uncertainties such as
the bremsstrahlung production rate (Sec. \ref{sec:SystBremDelta}).
Special attention was also given to avoid
the double counting of a systematic uncertainty as in the case of the momentum
resolution during the evaluation of the DC STRs (Sec. \ref{sec:SystSTRs}).
Table \ref{t:uncertainties} summarizes the systematic uncertainties by categories
that typically contain multiple independent uncertainties.

The weighted statistical uncertainties in Table \ref{t:uncertainties}
are computed from the statistical errors for the two
targets, weighted according to the target dependent systematic errors.
The weighted systematic uncertainties are the
quadrature sum of the target independent systematic
uncertainties and the appropriately weighted target
dependent systematic uncertainties.

As described above, we calculate the systematic uncertainties for $\rho$ and $\delta$
due to a mismatch $s$ as $(d\rho/ds)\sigma_s$ and $(d\delta/ds)\sigma_s$,
where $\sigma_s$ is our estimate of the possible size of the mismatch.
$\sigma_s$ is common to the $\rho$ and $\delta$ systematics,
so $(d\rho/ds)(d\delta/ds)\sigma_s^2$ represents a contribution
to the correlation between $\rho$ and $\delta$.
The correlation for the Ag (Al) target measurement is given by the sum of
the Ag (Al) and target independent correlations normalized by the 
quadratic sum of the Ag (Al) and target independent systematic uncertainties.
The final total correlation is the sum of the Ag and Al target correlations
weighted by the statistical weights used to determine the final decay parameter
measurement.

\begin{table}[!hbt]
	\caption{Systematic and statistical uncertainties for the $\rho$ and $\delta$ decay parameters.
	Most of the categories shown here are combinations of several independent uncertainties.}
	\label{t:uncertainties}
	\begin{tabular*}{1.0\columnwidth}{@{\extracolsep{\fill}}lcc}
	\hline\hline
	Category								& \multicolumn{2}{c}{Uncertainty ($\times 10^{-4}$)} \\
											& $\rho$ & $\delta$ \\
	\hline
	Target independent \\
	\quad Radiative corrections and $\eta$		& 1.3	& 0.6		\\
	\quad Momentum calibration                	& 1.2	& 1.2		\\
	\quad Chamber response						& 1.0	& 1.8		\\
	\quad Resolution                          	& 0.6	& 0.7		\\
	\quad Positron interactions\footnotemark[1]	& 0.5	& 0.1		\\
	\quad Others								& 0.3	& 0.4		\\
	Ag target\\
	\quad Bremsstrahlung rate				& 1.8	& 1.6		\\
	\quad Stopping position					& 2.0	& 6.0		\\
	\quad Target thickness					& 3.2	& 2.2		\\
	\quad Statistical						& 1.2	& 2.1		\\
	Al target\\
	\quad Bremsstrahlung rate				& 0.7	& 0.7		\\
	\quad Stopping position					& 0.2	& 0.8		\\
	\quad Statistical						& 1.3	& 2.4		\\
	Weighted systematic uncertainty			& 2.3	& 2.7		\\
	Weighted statistical uncertainty		& 1.2 	& 2.1 		\\
	Total uncertainty						& 2.6	& 3.4		\\
	\hline\hline
	\end{tabular*}
	\footnotetext[1]{excluding bremsstrahlung}
\end{table}

\subsection{Target independent uncertainties}
Two systematic uncertainties are external to the TWIST measurement. The uncertainty 
on the radiative corrections is given by the effect of the 
missing leading term $\mathcal{O}(\alpha^2)$ on the decay parameters \cite{Anastasiou}.
A numerical integration 
of this term in the TWIST fiducial regions showed that 
it has a similar shape, 5 times smaller than the $\mathcal{O}(\alpha^2 L)$ term. 
The spectrum shape of the $\mathcal{O}(\alpha^2 L)$ term is used to 
evaluate the change in decay parameters which gave a systematic uncertainty 
of $\pm0.16\times 10^{-4}$ ($\pm0.63\times 10^{-4}$) for $\rho$ ($\delta$). 
The second external uncertainty is due to the significant correlation factor
of 0.94 between the 
$\rho$ and the $\eta$ parameters.
The impact of this correlation on the decay parameters is evaluated 
by performing the decay spectra fit with $\eta$ fixed at the 
world average value lowered or raised by 1 standard deviation. 
The changes in decay parameters are used as systematic uncertainties and 
are equal to $\pm1.05\times 10^{-4}$ ($\pm0.12\times 10^{-4}$) for $\rho$ ($\delta$).

The chamber response category contains 
the systematic uncertainties for the STRs and the cathode foil position presented below,
and also the
asymmetry between upstream and downstream efficiency, the 
crosstalk, and the wire time offsets. The upstream-downstream asymmetry
uncertainty is measured by scaling the upstream half of the 
$p$-$\theta$ spectrum with respect to the downstream half
according to the difference in inefficiencies between data and simulation
extracted from the far upstream stops data 
(Sec. \ref{s:US-DS inef}). The corresponding systematic uncertainty for $\rho$ ($\delta$) is 
$\pm0.20\times 10^{-4}$ ($\pm0.75\times 10^{-4}$).
All crosstalk in the electronics of nearby wires in the drift 
chambers should be removed by the analysis software. An upper limit on a
potential systematic uncertainty due to remaining crosstalk 
is obtained by disabling the crosstalk removal and 
using the full change of $0.50\times 10^{-4}$ ($0.10\times 10^{-4}$) for $\rho$ 
($\delta$) as the uncertainty.
The wire time offsets are measured using different 
scintillators for the upstream and downstream halves of the detector which 
can lead to an asymmetry. The potential difference in this asymmetry 
between data and simulation (which was also calibrated)
is responsible for a systematic uncertainty of 
$\pm0.09\times 10^{-4}$ ($\pm0.44\times 10^{-4}$) for $\rho$ ($\delta$).

The spectrometer's reconstruction resolution in angle and
momentum is obtained from the far upstream stops data (Sec. \ref{sec:MC_validation}).
The $p$-$\theta$ spectrum is smeared on an event-by-event basis to exaggerate the effect
of a resolution mismatch.
The rescaled sensitivity provides a systematic uncertainty for the momentum resolution of
$\pm0.56\times 10^{-4}$ ($\pm0.70\times 10^{-4}$) for $\rho$ ($\delta$). The 
angle resolution mismatch leads to a systematic uncertainty $\lesssim0.1\times 10^{-4}$ for both parameters.

The positron interaction category (Table \ref{t:uncertainties}) includes
the systematic uncertainty for the backscattering of decay positrons
from outside material that
adds confusion to the track reconstruction.
The rate of backscattering positrons normalized to the muons stopping in the 
target is used to measure the mismatch in outside material between data and simulation.
The systematic effect of the outside material is determined by comparing
a nominal simulation and the simulation matching set 83 in which
the downstream beam package is added (Table \ref{t:data_sets_Al}).
The corresponding systematic uncertainty for $\rho$ ($\delta$) is $\pm0.48\times 
10^{-4}$ ($\pm0.13\times 10^{-4}$). The decay parameter difference between sets 83 and 84,
respectively with and without downstream beam package, is consistent with this uncertainty.

The category of systematic uncertainties under the name ``others'' in Table 
\ref{t:uncertainties} contains uncertainties that are $\lesssim 0.3\times 10^{-4}$
for both $\rho$ and $\delta$.
The overall spacing in $z$ of the wire chamber planes was established
to a fractional accuracy of $5\times10^{-5}$.
An analysis with the $z$ positions exaggerated by a fractional
change of $10^{-3}$ 
showed a corresponding change in the momentum calibration.
The changes in $\rho$ and $\delta$ were negligible.
A small correction to the magnetic field is obtained by fitting an analytic
function
to the difference between the measured field map
and the \textsc{Opera}-3D map. A comparison of set 84 analyzed with this correction
with the nominal analysis is used to obtain corrections and associated uncertainties.
This analysis shows significant corrections to the energy calibration parameters.
However, after the new calibration is applied, the change to $\rho$ and $\delta$
is $< 0.1\times 10^{-4}$.
Finally, the uncertainties for the muon and positron beam intensities are 
also part of this category and are negligible.

\subsection{Bremsstrahlung and $\delta$-electron production rate}
\label{sec:SystBremDelta}
A difference between data and simulation in the rates for 
the emission of bremsstrahlung photons or $\delta$ electrons would affect the decay parameter measurement.
Primarily these processes modify the positron momentum and angle between the muon decay vertex
and the beginning of the tracking region, thus altering
the reconstructed $p$-$\theta$\ spectrum shape.
Additionally a large change in positron
momentum within the tracking region can lead
to the identification of two separate track segments by the reconstruction algorithm.
This second effect 
reduces the reconstruction resolution by shortening the primary decay positron track,
but it can also be used to
compare the bremsstrahlung and $\delta$-electron production rates
for data and simulation.

\begin{figure}[tbh]
	\centering
	\includegraphics[width=\columnwidth]{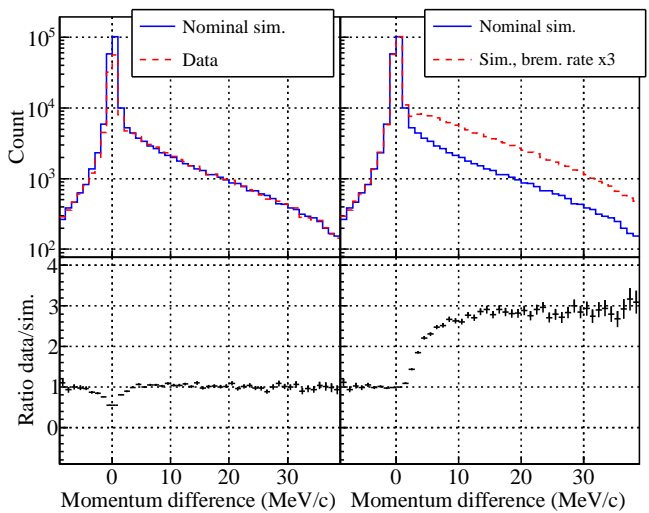}
	\caption{\label{fig:SystBrem}(color online) Number of events with two tracks
	versus the momentum difference between the two reconstructed tracks. The left-hand 
	side shows nominal data set 74 and the corresponding simulation.
	The right side shows nominal and exaggerated simulations. 
	The bottom plots correspond to the ratio of the two distributions above.}
	\label{f:Brem_BrknTrks}
\end{figure}

The bremsstrahlung production rate is evaluated by counting the number of events containing
two reconstructed tracks from a single 
decay positron. The data and simulation counts are normalized to the number of muons
stopping in the target.
The momentum of the bremsstrahlung photon is deduced from the momentum difference
$(\Delta p)$ between the two tracks and is shown in the left panel of
Fig. \ref{fig:SystBrem}.
The agreement between data and simulation is excellent except near $\Delta p=0$.
The discrepancy there could be due to the loss of hits from corners of
the drift cells, which happens more in data than in the simulation.
These additional hits lead to a higher rate of broken tracks with
very little momentum difference between the two track segments in the simulation.
Events with a $\Delta p$ between 15 and 35 MeV/$c$ are used for the comparison.
The average ratio of the bremsstrahlung production rates from all the data sets
to their corresponding simulations is equal to $1.024 \pm 0.004$.
Although this ratio is measured for the relatively low Z materials of the chambers,
it is assumed to be applicable for the full range of materials in the detector.
This assumption is supported by the target energy loss measurements
(Sec. \ref{sec:MC_validation}).
The bremsstrahlung rate is strongly dependent on the target material.
Thus the sensitivities to the production rate
are measured separately for each target,
from the difference in decay parameters between a nominal simulation
and a simulation with the bremsstrahlung production rate exaggerated by a factor of 3.
See the right panel of Fig. \ref{fig:SystBrem}.
The systematic uncertainties (Table \ref{t:uncertainties})
are given by the sensitivities rescaled by the factor $(3-1)/(1.024-1)=83.3$.
A simulation with a smaller exaggeration factor of 2 was also generated and analyzed,
and its results confirm the assumption of linearity of the systematic uncertainty.

Evaluation of the $\delta$ electron production rate uncertainty is done similarly to that of the bremsstrahlung. 
The production rate is measured by requiring a third track
from a negatively charged particle along with the two track segments from the decay positron. 
The momentum of the $\delta$ electron is measured directly from the reconstruction of the negatively charged track and used to 
select events with $\delta$ electrons in the momentum range ($6<p<16$) MeV/$c$.
The average ratio from all the data sets and 
their corresponding simulations is equal to $1.007 \pm0.009$.
The sensitivities to the $\delta$ electron production rate are also evaluated using a 
simulation with a threefold exaggerated rate.
The contribution is $\pm0.07\times10^{-4}$
($\pm0.06\times10^{-4}$) for $\rho$ ($\delta$) to the
Table \ref{t:uncertainties} positron interaction uncertainties.

\subsection{Momentum calibration}
\label{sec:SystEnergyCalib}
\subsubsection{End points fits}
\label{sec:SystFitECal}

The momentum mismatch between data and simulation at the end
point, which is assumed to be linear with respect to
$1/\!\cos\theta$ based on geometrical considerations, is
characterized by the parameters $a_{up}, b_{up}, a_{dn}$, and
$b_{dn}$ as shown in Eq. (10). 
However, if one assumes that the uncertainties are purely
statistical, the linear fits result in a total $\chi^2$ of 212.9
for 168 degrees of freedom, corresponding to a $p$ value from all
the data sets equal to 0.011.  Evidently the behavior
of the mismatch is not linear, possibly due to higher order
effects or perhaps some underlying fine structure in the momentum
spectrum for each angle bin.  The manifest nonlinearity stems
from the upstream end-point portions of the fits, while the
downstream portions have larger statistical uncertainties that
could mask any nonlinear behavior.

To account for this nonlinearity we add in quadrature an
uncertainty of 1.6 keV/$c$ to the statistical uncertainty of the
momentum mismatch at each $1/\!\cos\theta$ bin, in order to
achieve an upstream reduced $\chi^2$ of one. When propagated to
the uncertainties of the decay parameters, this results in
systematic uncertainties of $\pm0.58 \times 10^{-4}$ ($\pm0.54
\times 10^{-4}$) in $\rho$ ($\delta$).

The observed offset at the end-point between data and simulation
is $\sim$10 keV/$c$.  To understand this difference quantitatively,
a number of sources of systematic corrections and uncertainties
must be considered.  Approximately 4 keV/$c$ of this total is due
to a slightly incorrect scale used for the magnetic field in the
simulation.
Small corrections to the momentum calibration parameters were
calculated from the best values for the target thicknesses,
magnetic field map, and match of the muon stopping distribution
(to be described in Sec. \ref{sec:PostBlindSyst}). Systematic
uncertainties for these parameters have been determined using the
same simulation studies that were used to determine the muon
decay parameter uncertainties.  The most significant items are
from the magnetic field map, the STRs, the $z$ spacing of the
chambers, the match of the muon stopping distribution, and the
target thicknesses.  Additionally, there is an uncertainty from
soft momentum loss in the simulation, consisting of an ionization
momentum loss uncertainty of 2\% and a radiative momentum loss
uncertainty of 3\% \cite{Berger} for the drift chamber and target
materials used.  After corrections, the magnitude of the mean
slope parameters for each target is less than 5 keV/$c$, and the
mean offset magnitude is less than 7 keV/$c$, both with
systematic errors of $\approx$5 keV/$c$. This level of agreement
shows acceptable consistency of the data with the expected
accuracy of the simulation.

\subsubsection{Propagation model}
\label{sec:SystPropagECal}
The momentum mismatch between data and simulation is measured only at
the kinematic end-point but is corrected over the entire spectrum. The
predicted momentum dependence of this calibration depends on the source of the
momentum mismatch between data and simulation. For instance, a difference
in solenoid magnetic field strength leads to a momentum mismatch that
depends linearly on the momentum and is referred to as a \textit{scale}.
Another example is a mismatch in target thickness, which translates into
an angle-dependent \textit{shift} of the momentum (to first order), with
the angle dependence measured by the slopes $a_{up}$ and $a_{dn}$. Most of the
observed offset at the end-point could not be attributed to
a unique source. Therefore it was assumed that the
propagation of the momentum mismatch is a mixture of \textit{shift} and
\textit{scale}.

For this reason the decay parameters were computed for the two extreme
cases of propagation which correspond to a pure \textit{shift} with the form
\begin{linenomath}
\begin{equation}
	p_{\mbox{\,corrected}} = p_{\mbox{\,reconstructed}} - \left( b - \frac{a}{|\cos\theta|}\right),
\end{equation}
\end{linenomath}
or a pure \textit{scale}, given by
\begin{linenomath}
\begin{equation}
	p_{\mbox{\,corrected}} = \frac{p_{\mbox{\,reconstructed}}}{ 1 + \frac{1}{\sqrt{W_{e\mu}^2 - m_e^2}} \big( b - \frac{a}{|\cos\theta|}\big)}
\end{equation}
\end{linenomath}
where $W_{e\mu}$ and $m_e$ were defined in the context of
Eq. \ref{eq:mudecay_michel}.
The average values of the $\rho$ and $\delta$ parameters using 
the \textit{shift} and the \textit{scale} propagations are different
respectively by $2.04\times 10^{-4}$ and $2.16\times 10^{-4}$.
Their mean is used for the decay parameter.
Half of the difference between \textit{shift} and \textit{scale} 
is used as the uncertainty to cover the two extreme possibilities.
Therefore the systematic uncertainty from the propagation model is 
$\pm1.02\times 10^{-4}$ ($\pm1.08\times 10^{-4}$) for $\rho$ ($\delta$).

\subsection{DC STRs}
\label{sec:SystSTRs}
The accuracy of the helix reconstruction depends on the quality of the STRs.
In particular, differences between the respective accuracies of
data and simulation STRs can lead to a bias in the decay parameter measurement.
The STRs were derived in both cases using 
the two-dimensional time residual distributions covering the entire drift cell ($T_{res}$) from the helix fitter.
The sensitivity to a mismatch in STRs is 
measured by creating simulation STRs containing the difference $\Delta T_{res}$ between data and simulation STRs.
First 44 $\Delta T_{res}$ 
are created by taking the difference between the data and the simulation $T_{res}$ for each DC. The 44 $\Delta T_{res}$
are fitted with a fifth order polynomial function to guarantee the smoothness of the STRs created in the 
next step. Second 44 STR tables are created by adding the 44 polynomial functions exaggerated by a factor of ten 
to 44 duplicates of the simulation STRs.
A simulation is reanalyzed with these new STRs and this set is fitted against the unmodified set to measure 
a change in decay parameters.
The corresponding sensitivity of the decay parameters
to the STRs changes significantly if the 
propagation model for the momentum calibration is a \textit{shift} or a \textit{scale}.
For this reason the sensitivities from both 
models are averaged and the total sensitivity is $-7.5 \times 10^{-4}$ ($-14.5 \times 10^{-4}$)
for $\rho$ ($\delta$) to STRs exaggerated by a factor of 10.

The momentum resolution at the kinematic end-point is very different between the standard and these exaggerated-STR analyses of the 
simulation. However the impact of the resolution on the decay parameters is already taken into account in a separate 
systematic uncertainty. The systematic effect from the resolution must be subtracted from the STR sensitivities
evaluated in this section to avoid double counting.

The sensitivities to the reconstruction resolution are evaluated from the differences
in decay parameters between a nominal spectrum and spectra created
with the events smeared in momentum by different values.
This procedure is equivalent to
a degradation in resolution.
The contributions of the resolution to the STR sensitivities,
which must be subtracted from the total 
sensitivities, are $-4.4\times 10^{-4}$ and $-4.5\times 10^{-4}$ for $\rho$ and $\delta$.
Finally each sensitivity is scaled down by the 
exaggeration factor of 10 to give the systematic uncertainty
for the DC STRs of $\pm0.3\times 10^{-4}$ ($\pm1.0\times 10^{-4}$) 
for $\rho$ ($\delta$).

\subsection{Cathode foil position}
\label{sec:SystFoilPos}
The relative position of each cathode foil with
respect to adjacent anode wires has
two effects on the detector response.
First of all it modifies the electric field and consequently the STRs.
This effect is included in the 
plane dependent measurement of the STRs in data and therefore
does not lead to any additional systematic uncertainty.
The second effect is the change of the drift cell size
which can change the number of cells crossed by each positron.
This has an impact on the track reconstruction, in 
particular on the resolution of the left-right ambiguity in the helix fitter.

In the apparatus there are two different sources of uncertainty on the foil position. The 
first uncertainty is due to the DC outer foil bulging
toward or away from the wires as a result of the differential pressure between
the DCs and the helium-nitrogen gas mixture surrounding the chambers.
The permanent foil bulge was toward the wires (60$\pm$22) $\mu$m (average)
during the 2006 run period, and away from the wires (8$\pm$22) $\mu$m during 2007. 
The second source of uncertainty comes from the construction of
the chambers and was estimated to be $\pm$100 $\mu$m on average. 

The sensitivity of the decay parameters to the cathode foil position is evaluated by generating a simulation with cathode 
foils moved toward the wires by 500 $\mu$m, without modifying the STRs.
The fit of this exaggerated simulation against the corresponding nominal 
simulation gives a sensitivity of $4.0\times 10^{-4}$ and $5.9\times 10^{-4}$ for $\rho$ and $\delta$.
In this modified simulation the drift cell size is reduced
for all the planes but in reality some drift 
cells are potentially larger in data than they are in simulation.
Therefore the average systematic effect is smaller than
the estimated sensitivity.
For this reason the cathode foil position uncertainties
from the bulge and the chamber construction are not added in quadrature 
but instead only the largest uncertainty of $\pm$100 $\mu$m is considered, leading to an exaggeration factor of 5.
The corresponding rescaled systematic uncertainty is $\pm0.80\times10^{-4}$ 
($\pm1.18\times10^{-4}$) for $\rho$ ($\delta$).

\subsection{Statistics bias correction}

A sensitivity to the difference in statistics between data and simulation 
was discovered in the $\chi^2$ minimization technique used by the decay 
parameter fit and the momentum calibration. In the situation where the 
data and simulation spectra have the same number of events, the 
difference of the two asymmetric Poisson distributions of two bins leads 
to a symmetric probability distribution for the residuals. However all the 
simulations contain 2 to 3 times more events than their corresponding data set 
to reduce the statistical uncertainty for the decay parameters. This creates 
an asymmetric distribution for the residuals and a bias
in the $\chi^2$ minimization \cite{Barlow}.

The biases of the decay parameter fit and the momentum calibration
fit were evaluated by performing the fits
between a data set 
and subsets of the simulation with matching statistics. For each fit 
parameter, the difference between the average of the subsets
and the results using 
the whole simulation corresponds to the bias.
Corrections of $-0.20\times10^{-4}$ and $-0.05\times 10^{-4}$ for $\rho$ and $\delta$
were applied
to account for the average fitting bias of the
decay parameters. The momentum calibration fitting bias
corrections were applied differently, on a set by set basis, and on
the decay parameter measurements from the 
\textit{shift} and the \textit{scale} propagation of the calibration to the spectrum.
The values of the corrections range between $-0.92\times 10^{-4}$ and
$-1.36\times 10^{-4}$ ($-0.31\times 10^{-4}$ and $-0.53\times 10^{-4}$)
for $\rho$ ($\delta$).

\section{Postblind analysis}
The hidden parameters of the blind analysis were revealed once
the differences of the three decay parameters between data and
simulation were confirmed and the systematic uncertainties were
fully evaluated. The results for the three parameters were
consistent with the SM.
However, the product
$P_\mu\xi\delta/\rho$ was $1.001\,92 \ ^{+0.001\,67}_{-0.000\,66}$.
Although the sign of the deviations of the individual decay
parameters from the SM is not constrained in the generalized
matrix element treatment \cite{Fetscher:1986}, the product
$P_\mu\xi\delta/\rho$ must be $\le 1$. This product can be
identified with the asymmetry between the extremes of $\cos\theta
= \pm1$ and evaluated at $x = 1$ by using
Eqs. \eqref{eq:mudecay_michel}, \eqref{e:Fisotropic}, and \eqref{e:Fanisotropic}.

A measurement of $P_\mu\xi\delta/\rho > 1$ could have
been due to the matrix element treatment or the momentum-angle
functional form being inadequate to describe the data, but it
could also have been due to a systematic uncertainty or
correction missing or not evaluated properly in the analysis.
Furthermore, $P_\mu\xi\delta/\rho$ from the blind analysis was
different for the Ag and Al target data by 3.8 $\sigma$. Both the
large value of $P_\mu\xi\delta/\rho$ and the mismatch between Ag
and Al target data triggered an exhaustive review of the blind
analysis and special scrutiny of various systematic effects that
could explain these results.
Among the tests performed, effects such as $\mu^+ \rightarrow e^+ 
X^0$\ decays (where $X^0$\ is a long-lived unobserved particle), an incorrect 
value of the $\eta$\ parameter, or plausible errors in the radiative 
correction implementation, did not resolve the mismatch. However we found two 
corrections that were missed during the blind analysis.

\subsection{Additional systematic uncertainties and corrections}
\label{sec:PostBlindSyst}
The effect of the muon radiative decay on the $p$-$\theta$ spectrum 
is included in the radiative corrections
of the decay positron spectrum of the simulation (Sec. \ref{s:simulation}).
However we neglected to simulate the photon from radiative decays.
Although the wire chambers are 
insensitive to photons, the electrons and positrons from pair production 
or Compton scattering of the photons can affect the track reconstruction. 
These processes occur at different rates in Ag and Al and therefore potentially bias
the Ag and Al measurements differently.
The effect on the decay parameters is measured using two simulations 
of pure muon radiative decay using the Fronsdal and \"Uberall formula 
\cite{fron59} to calculate the momentum and angle of the decay positrons 
and the photons. One simulation contains all the standard physics processes 
while the second simulation does not include the pair production and 
the Compton scattering of the photon so that, as in the nominal
simulation, the radiative decay photons are absent.
The decay parameter difference is renormalized 
using the branching ratio of $(1.4\pm0.4)$\% from \cite{PDG}.
The corrections to $\rho$ and $\delta$ for the Ag data 
are $0.59\times10^{-4}$ and $0.76\times10^{-4}$, and they are negligible for the Al data.

The second category of corrections and related refinement of systematic uncertainties
is due to the large sensitivity to energy loss
through bremsstrahlung emission by the positron as it travels through the target.
We expected the momentum calibration to correct for a
mean muon stopping position difference (MSPD) between data and simulation.
A match with precision at the level of 1 $\mu$m is required, but
the kinematic edge was not very sensitive to the large changes in momentum
due to bremsstrahlung that can affect the spectrum in the fiducial region.
Also, we assumed that $a_{up} = a_{dn} = 0$
corresponded to MSPD $= 0$, which turned out not to be true.
An improved technique was developed to determine MSPD.
For the blind analysis the systematic uncertainty for a mismatch in bremsstrahlung production rate
was evaluated only for the Ag target and applied to all the data.
In the postblind procedure, it is
evaluated for each target, and a separate uncertainty is added for
a mismatch in target thickness.

\begin{figure}[bt]
	\centering
	\includegraphics[width=\columnwidth]{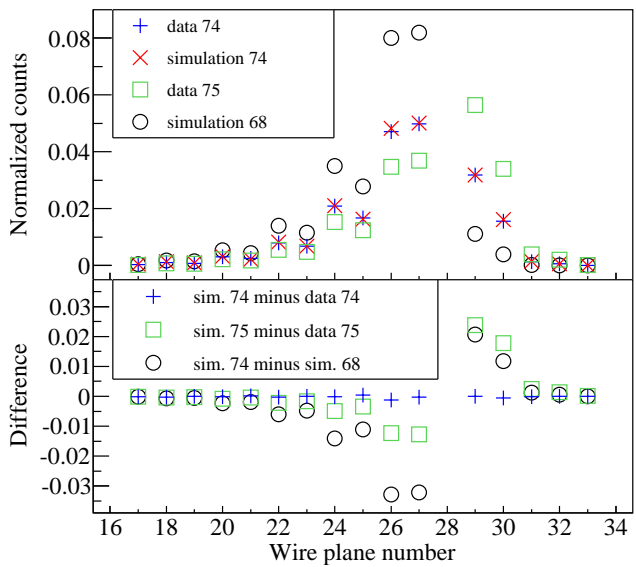}
	\caption{
	(color online) Distributions of number of muon tracks ending in a wire plane,
	versus plane
	number (top panel), normalized to number ending in PC 6 (plane
	number 28), for different data and simulation conditions. The
	bottom panel shows the differences of selected pairs of
	distributions.
	}
	\label{f:Frac}
\end{figure}

A measurement based on the distribution of the last wire plane 
hit by the muons is used to evaluate the muon's 
MSPD between data and simulation.
The last wire plane distributions are normalized to the number of 
muons stopping in the target defined by the counts in PC 
6 (wire plane 28) which is located just upstream of the target (Fig. \ref{f:Frac}).
The differences between data and simulation last wire plane distributions 
show agreement at the percent level for all planes for 
data sets 68, 74 and 76, which confirms a match 
in the mean and the widths of the stopping distributions.
However, disagreements for other sets identified a sensitivity to
MSPD. Different methods were tested to establish MSPD with
improved precision. It was found that MSPD could be measured
using an average of the PC 5 and PC 7 fractional differences
(wire planes 27 and 29) where the sensitivity is the highest.
We verified that MSPD measurements from other planes 
are consistent with the measurement from PC 5 and PC 7.
The relationship between the average fractional difference and the 
MSPD is extracted from the comparison of simulations with known
non-zero MSPDs such as between the simulations of the data sets 
68 and 74 shown Fig. \ref{f:Frac}. Each data set and its 
corresponding simulation were compared and MSPDs of up to 1.6 $\mu$m
were determined for the Ag target and 3.8 $\mu$m for the Al target.
The sensitivity of the decay parameters to MSPD is determined 
by creating $p$-$\theta$ spectra for different depth intervals in the target 
using the true stopping position of the muons in the simulation. 
Set by set corrections are applied and range from 0.0 to 
$-3.3\times10^{-4}$ (0.0 to $-9.8\times10^{-4}$) for $\rho$ ($\delta$).
Although MSPD is larger in Al, the largest corrections are for 
the Ag target (set 75) because of a higher density and bremsstrahlung production rate in Ag.
We estimate the MSPD uncertainty to be 1 $\mu$m for Ag and 
2 $\mu$m for Al. The systematic uncertainties on the correction,
determined from the sensitivity to MSPD, are respectively for the Ag and Al targets
$\pm 2.0\times10^{-4}$ ($\pm 6.0\times10^{-4}$)
and $\pm 0.2\times10^{-4}$ ($\pm 0.8\times10^{-4}$)
for $\rho$ ($\delta$) (under ``stopping position'' in Table \ref{t:uncertainties}).

The accurate measurement of target thickness is based on a
destructive test that could only be performed after the
experiment. It gives a thickness of $(30.9 \pm 0.6)\ \mu$m
and $(71.6 \pm 0.5)\ \mu$m for the Ag and Al targets respectively.
The simulation used the prior estimate of
the Ag (Al) target thickness of 29.5 $\mu$m (71.0 $\mu$m), which
was based on measurements of
material samples that were similar to the targets, but obviously
not identical. The impact 
on the decay parameters of the mismatch in Ag target thickness 
was determined by generating a simulation with a 65 $\mu$m thick 
target. The corresponding systematic uncertainty for $\rho$ ($\delta$) is $\pm 3.2\times10^{-4}$ 
($\pm 2.2\times10^{-4}$). On the other hand, the systematic uncertainties for the 
mismatch in Al target thickness were negligible with values $ < 0.3\times10^{-4}$.

\subsection{Results}
\begin{figure}[bt]
	\includegraphics[width=\columnwidth]{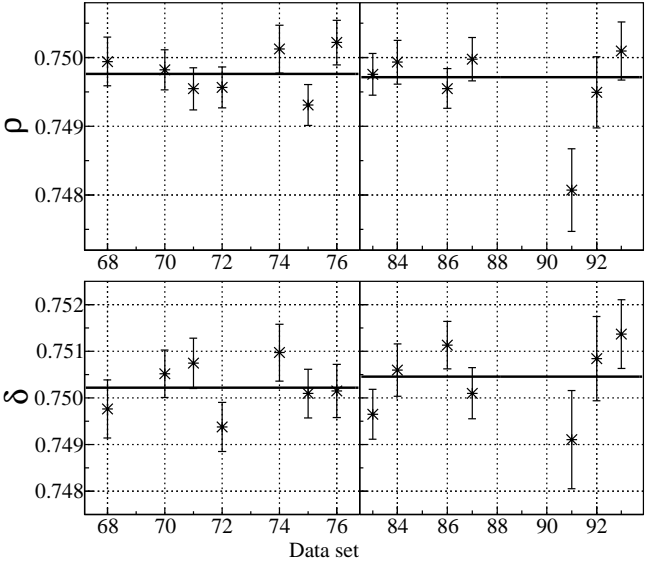}
  \caption{Results of the $\rho$ and $\delta$
  measurement for each data set, fitted separately
  for the Ag sets (left panels) and
  for the Al sets (right panels). Only statistical
  uncertainties are shown here.}
    \label{f:Consistency}
\end{figure}

The final results are extracted from all data sets identified
as valid for analysis of $\rho$ and $\delta$.
These sets are unchanged from the
blind analysis. However, the postblind result for $P^\pi_\mu\xi$ includes
correlation information from the measurement of $\delta$ in five data sets
not used for $P^\pi_\mu\xi$, which reduces its statistical uncertainty but does
not change the central value.
This analysis has already been published separately \cite{Jamespmuxi}.
The consistency is shown in Fig. \ref{f:Consistency} for
measurements taken under various experimental conditions
(Tables \ref{t:data_sets_Ag} and \ref{t:data_sets_Al}), demonstrating that the simulation
reproduces these conditions accurately.
The Ag and Al data are fitted separately and then are combined using the
target dependent systematic uncertainties.

The additional corrections and systematic uncertainties determined
during the reevaluation of the analysis (Sec. \ref{sec:PostBlindSyst})
change the central values of $\rho$ and $\delta$ by 
$-1.4\times 10^{-4}$ and $-2.3\times 10^{-4}$;
both changes are less than the total assessed systematic
uncertainties, which themselves changed by less than $0.6\times 10^{-4}$. 
The modified 
values of $P^\pi_\mu\xi\delta/\rho$ for Ag and Al data are now consistent
within $\approx 1 \sigma$, while
$P^\pi_\mu\xi\delta/\rho = 1.001\,79 \ ^{+0.001\,56}_{-0.000\,71}$ has
decreased but remains somewhat greater than unity.
The final TWIST 
results for $\rho$ and $\delta$ are
\begin{linenomath}
$$\rho = 0.749\,77\pm0.000\,12\mbox{(stat.)}\pm0.000\,23\mbox{(syst.)};$$
$$\delta = 0.750\,49\pm0.000\,21\mbox{(stat.)}\pm0.000\,27\mbox{(syst.)};$$
\end{linenomath}

\begin{figure}[!hbt]
	\includegraphics[width=3.4in]{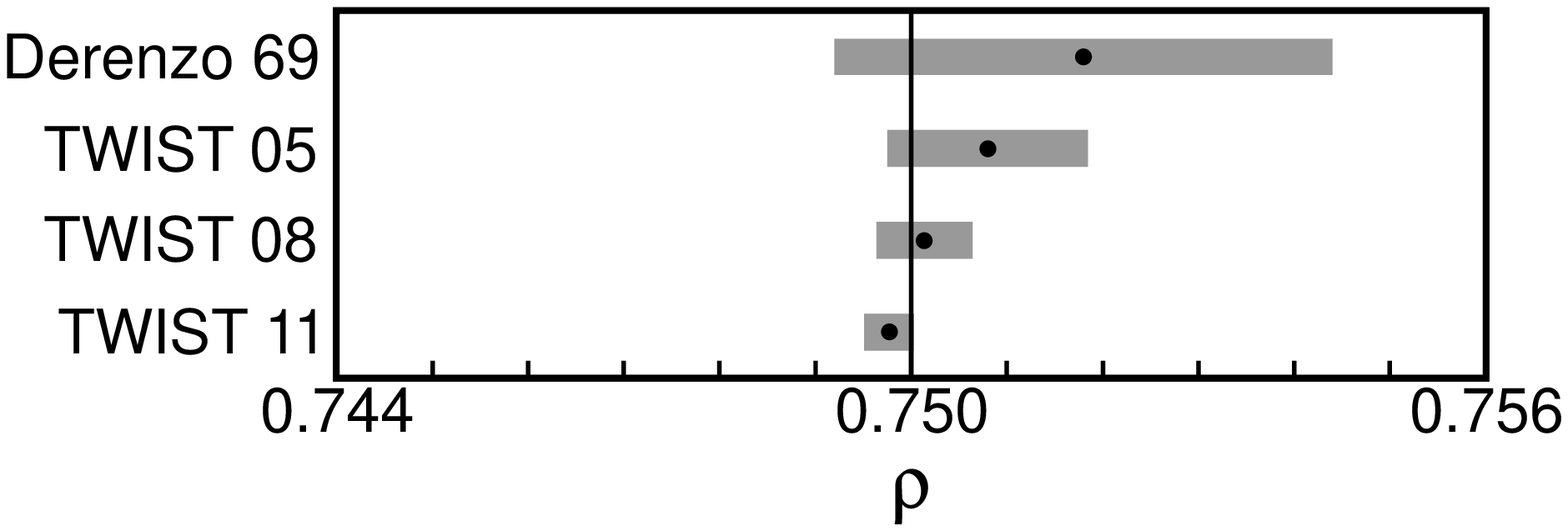}
	\includegraphics[width=3.4in]{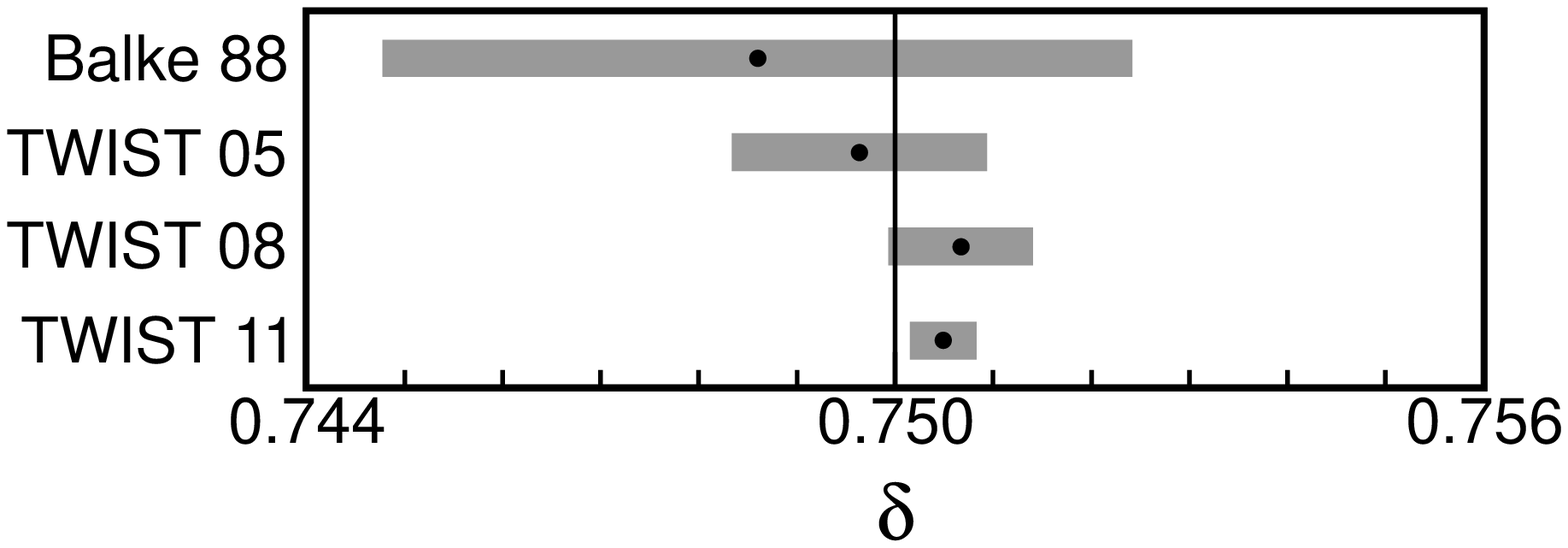}
    \caption{Summary of the central values and total uncertainties of $\rho$ 
	and $\delta$ from this analysis, along with the previous
	published measurements \cite{musser:2005,gaponenko:2005,Robpaper,Derenzo,Balke}.}
    \label{f:VersusPrev}
\end{figure}

These results represent an improvement of a factor of respectively 14 
and 11 over the pre-TWIST direct measurements. They are consistent 
with the SM predictions of $\rho = \delta = 0.75$ 
and furthermore agree with previous measurements (Fig. \ref{f:VersusPrev}).

\section{Theoretical implications}

\subsection{Global Analysis of Muon Decay}
\label{sec:globalanalysis}

A new global analysis of all available muon decay data has been performed 
including the final TWIST results for the decay parameters and their 
correlations \cite{Jamespmuxi}. All other input values are the same as in the
analysis of ~\cite{gagliardi:2005}. The global analysis used 
a Monte Carlo method similar to that of ~\cite{burkard:1985} 
to map out the joint probability distributions for 10 variables 
(see Table~\ref{tab:newglobalanalysis}), each of which is a bilinear
combination of the weak coupling constants $g_{\epsilon\mu}^\gamma$. 
The constraint of $Q_{RR} + Q_{RL} + Q_{LR} + Q_{LL} = 1$
is applied (see Eq. \eqref{eq:Qem}), 
resulting in 9 independent variables; the best fit values and 90\% 
confidence limits are given in Table~\ref{tab:newglobalanalysis}.
The decay parameters could then be written in terms of these independent 
variables, and the results are included in Table~\ref{tab:newglobalanalysis}.
The present analysis makes significant improvements in the limits on 
$Q_{RR}$, $Q_{LR}$, and $B_{LR}$
compared to the 2005 analysis, and tightens several of the other limits.  

\begin{table}[bt]
  \caption{Results of a new global analysis of muon decay data,
  including the present measurements (parameter definitions in \cite{gagliardi:2005}).
  Best fit values and 90\% confidence limits are given.
  $P_\mu^\pi=1$ is assumed. Global analysis values of the decay
  parameters are also listed.}
  \label{tab:newglobalanalysis}
  \begin{tabular*}{1.0\columnwidth}{@{\extracolsep{\fill}}ll}
    \hline
    \hline
     Parameter & Global analysis results (\e{-3}) \\
    \hline
    $Q_{RR}$ & $ <0.30 \, (0.16 \pm 0.11)$ \\
    $Q_{LR}$ & $ <0.63 \, (0.39 \pm 0.18)$ \\
    $Q_{RL}$ & $ <44 \, (25 \pm 13)$ \\
    $Q_{LL}$ & $ >955 \, (974 \pm 13)$ \\
    $B_{RL}$ & $ <11 \, (6.5 \pm 3.3)$ \\
    $B_{LR}$ & $ <0.52 \, (0.30 \pm 0.15)$ \\
    $\alpha/A$ & $ 0.1 \pm 1.4$ \\
    $\beta/A$ & $ 1.4 \pm 2.4$ \\
    $\alpha'/A$ & $ -0.1 \pm 1.4$ \\
    $\beta'/A$ & $ -0.5 \pm 2.4$ \\
    \hline
     Parameter & Global analysis results \\
    \hline
    $\rho$ & $ 0.749\,60\pm 0.00019$ \\
    $\delta$ & $ 0.749\,97\pm 0.000\,28$ \\
    $\xi$ & $ 0.998\,97\pm 0.000\,46$ \\
    $\eta$ & $ -0.002\,7\pm 0.005\,0$ \\
    \hline
    \hline
  \end{tabular*}
\end{table}

The results from this global analysis can be
used in Eq.~\eqref{eq:Qem} to place limits on the magnitudes of the
weak coupling constants $\abs{g_{\epsilon\mu}^\gamma}$; the exceptions
are $\abs{g_{LL}^V}$ and $\abs{g_{LL}^S}$, which are determined more
sensitively from inverse muon decay, {$e^-\nu_\mu \rightarrow \mu^-\nu_e$}.
These limits are presented in Table~\ref{tab:new_coupling_limits}.
Tighter limits from the present analysis of up to a factor of 2 compared
to the 2005 analysis are found in $|g^{\gamma}_{RR}|$ and $|g^{\gamma}_{LR}|$.

\begin{table}[tbh]
   \caption{Limits on the weak coupling constants.  (Limits on
     $|g_{LL}^S|$ and $|g_{LL}^V|$ are from Ref.~\cite{PDG}.)}
   \label{tab:new_coupling_limits}
   \begin{tabular*}{1.0\columnwidth}{@{\extracolsep{\fill}}lll}
   \hline
   \hline
    $|g_{RR}^S| < 0.035$ & $|g_{RR}^V| < 0.017$ & $|g_{RR}^T| \equiv 0$ \\
    $|g_{LR}^S| < 0.050$ & $|g_{LR}^V| < 0.023$ & $|g_{LR}^T| < 0.015$ \\
    $|g_{RL}^S| < 0.420$ & $|g_{RL}^V| < 0.105$ & $|g_{RL}^T| < 0.105$ \\
    $|g_{LL}^S| < 0.550$ & $|g_{LL}^V| > 0.960$ & $|g_{LL}^T| \equiv 0$ \\
    \hline
    \hline
  \end{tabular*}
\end{table}

A new indirect limit on the value of $P_{\mu}^{\pi}\xi\delta/\rho$
can be obtained from the global analysis. The linear combination
\begin{equation}
\rho - \xi\delta = \frac{3}{2}Q_{RR} + 2(Q_{LR}-B_{LR}),
\end{equation}
combined with the constraints $0 \le Q_{RR}$, $0 \le B_{LR} \le Q_{LR}$,
and $P_{\mu}^{\pi} \le 1$, builds in the physical condition
$P_{\mu}^{\pi}\xi\delta/\rho \le 1$, which is required
to avoid a negative muon decay probability near the end point
[see Eq. \eqref{eq:mudecay_michel}].
We find $P_{\mu}^{\pi}\xi\delta/\rho = 0.999\,47 \pm 0.000\,28$
or $P_{\mu}^{\pi}\xi\delta/\rho > 0.999\,09$ (90\% C.L.).
This is a significant improvement over the previous
limit of $P_{\mu}^{\pi}\xi\delta/\rho > 0.996\,82$ \cite{Jodidio}.

The quantity $Q_R^\mu = Q_{RR} + Q_{LR}$ represents the total
probability for a right-handed muon to decay into any type of
electron, a process forbidden under the SM weak
interaction.  The new limits on $Q_{RR}$ and $Q_{LR}$ shown in
Table~\ref{tab:newglobalanalysis} yield a new 90\% confidence
limit upper bound on the combined probability $Q_R^\mu<0.000\,82$, a
factor of 6 improvement over the limit from the
pre-TWIST numbers.

\subsection{Neutrino mixing}
It is now established that flavor mixing occurs in the neutrino states
\cite{SuperK, SNO}. Thus the neutrino state summations in
Eq. \eqref{eq:mudecay_michel} need to
extend over the additional kinematically allowed states and mixings
since the matrix elements of Eq. \eqref{eq:mudecay_matrixelem}
are evaluated in the flavor basis.
Doi \textit{et al.} \cite{Doi} have calculated the
dependence of the decay parameters
on the neutrino masses and mixings when only chiral vector weak
couplings are allowed. These calculations have shown that, when all of the
neutrinos involved are much lighter than the muon mass, as is the case
for $\nu_e$, $\nu_\mu$, and $\nu_\tau$, the decay parameters
in Eqs. \eqref{e:Fisotropic} and \eqref{e:Fanisotropic} are unaffected.
However if there are additional mixed neutrino states with
large right-handed Majorana masses, the decay rate is modified. For seesaw model
extensions of the SM, the effect on the muon decay parameters is below the
precision of the present measurement.

\subsection{Nonlocal tensor interaction}
\label{sec:NonLocalTensorInt}
The coupling constants $g_{RR}^T$ and $g_{LL}^T$ are set to zero in 
the general 4-fermion interaction [Eq. \eqref{eq:mudecay_matrixelem}]
because their corresponding matrix elements 
cancel out. However by abandoning locality \cite{Chizhov94, Chizhov_Review},
one can redefine the tensor interaction as
\begin{equation}
	\Gamma^T \otimes \Gamma^T = \frac{1}{2}\ \sigma^{\alpha\lambda} \otimes \sigma_{\beta\lambda} \cdot \frac{4q_\alpha q^\beta}{q^2}
\end{equation}
where $q_\mu$ is the momentum transfer of some virtual boson.
This form of the tensor interaction permits nonzero contributions from $g_{RR}^T$ and $g_{LL}^T$, in addition to $g_{RL}^T$ and $g_{LR}^T$.
If the tensor interaction couples equally to quarks and leptons,
pion decay data require $g_{LL}^T$, $g_{RL}^T$, and $g_{LR}^T$ to be very small \cite{Voloshin, Chizhov_Review}.
The seesaw mechanism to generate neutrino masses would require these same three coupling constants to be identically zero in muon decay.
For these reasons, Chizhov \cite{Chizhov_Review} explored how the muon decay spectrum would be
changed if the standard model were augmented by the addition of a single additional coupling constant, $g_{RR}^T$.

Nonzero $g_{RR}^T$ requires the introduction of
a new muon decay parameter, $\kappa\ (\approx g_{RR}^T)$,
in the differential muon decay spectrum, such 
that Eqs. \eqref{e:Fisotropic} and \eqref{e:Fanisotropic} become:
\begin{eqnarray}
  F_{\mathrm{IS}}(x) & = & x(1-x)
  + \frac{2}{9} \rho \left( 4x^2 - 3x - x_0^2 \right)\nonumber\\
  && + \eta x_0 (1-x) + \kappa x_0+ F_{\mathrm{IS}}^{\mathrm{RC}}(x), \label{e:Fiso2}
\end{eqnarray}
\begin{eqnarray}
  F_{\mathrm{AS}}(x) & = & \cfrac{1}{3} \xi\sqrt{x^2-x_0^2}
  \big[ 1 - x + \cfrac{2}{3}\,\delta \big( 4x - 3 \big.\big. \nonumber\\
  && \big.\big.+ \big( \sqrt{1-x_0^2} - 1 \big) \big) \big] + \kappa x_0 (2-x) \nonumber\\
  && + \xi F_{\mathrm{AS}}^{\mathrm{RC}} (x).\label{e:Faniso2}
\end{eqnarray}
The $\kappa$ term in Eq. \eqref{e:Fiso2} introduces a negligible distortion
in the isotropic distribution relative to the precision of TWIST.
In contrast, the linear $\kappa$ term in Eq. \eqref{e:Faniso2} represents a significant modification to $F_{\mathrm{AS}}$.
Chizhov \cite{Chizhov_Review} finds that it increases the integral forward-backward asymmetry by a factor $18 x_0 \kappa$.

The three decay parameters measured by TWIST also receive direct contributions from $g_{RR}^T$:
\begin{eqnarray}
	\rho = \cfrac{3}{4} ( 1 - 2 \kappa^2 ),\qquad & \xi = 1 + 2\kappa^2,\nonumber\\
	\xi \delta = \cfrac{3}{4} ( 1 - 4 \kappa^2 ),\qquad & \delta = \cfrac{3}{4} ( 1 - 6 \kappa^2 ) .
\label{e:kappa_quad}
\end{eqnarray}
Thus, $\kappa$ provides both linear and quadratic modifications to the muon decay spectrum.
This combination implies the linear fitting procedure used in this analysis
[Eq. \eqref{eq:DiffDataMCSpectra}] cannot be altered to fit $\kappa$ directly.

A study \cite{Williams} was performed using the theoretical $p$-$\theta$ spectrum to determine how a nonzero value
of $\kappa$ would distort the values we obtain for $\rho$, $\delta$, and $P_{\mu}\xi$.
We find
\begin{eqnarray}
\rho_{eff} & \approx & \frac{3}{4} (1 - 0.4 x_0 \kappa - 2 \kappa^2) , \nonumber \\
(P_{\mu}\xi)_{eff} & \approx & 1 + 16.5 x_0 \kappa + 2 \kappa^2 , \nonumber \\
(P_{\mu}\xi\delta)_{eff} & \approx & \frac{3}{4} (1 - 1.2 x_0 \kappa - 4 \kappa^2) .
\label{e:kappa_eff1}
\end{eqnarray}
The latter two equations imply
\begin{equation}
\delta_{eff} = \frac{(P_{\mu}\xi\delta)_{eff}}{(P_{\mu}\xi)_{eff}}
\approx \frac{3}{4} (1 - 17.7 x_0 \kappa - 6 \kappa^2) .
\label{e:kappa_eff2}
\end{equation}
When Eqs.\@ (\ref{e:kappa_eff1}) and (\ref{e:kappa_eff2}) are combined with our measured values for $\rho$, $\delta$, $P_{\mu}\xi$,
and their correlations to calculate the probability distribution for $\kappa$, we find $-0.009 < \kappa < +0.000\,5$ (90\% C.L.).

\section{Summary}
\label{sec:conclusion}

These new measurements of the muon decay spectrum culminate the TWIST
experimental program and are about 1 order of magnitude
more precise for each one of the three decay
parameters than measurements prior to TWIST. 
In fact, it has been more than 40 years
since the previous precision measurement of $\rho$, and no experimental effort
in the intervening years has succeeded in surpassing the precision quoted in Ref. 
\cite{Derenzo}, until TWIST. For $\delta$, the interval has been more than 20
years since the last measurement \cite{Balke}. Our final results supersede
our intermediate values.
They are consistent with SM predictions,
placing more stringent limits on physics beyond the
SM in the weak interaction.

\section*{Acknowledgments}
We thank all early TWIST collaborators and students
for their substantial contributions. 
We also thank C.~Ballard, M.~Goyette, S.~Chan, A.~Rose,
P.~Winslow, and the
TRIUMF cyclotron operations, beamlines, and support personnel.
Computing resources were provided by WestGrid and Compute/Calcul Canada.
This work was supported in part by the Natural Sciences and Engineering 
Research Council and the National Research Council of Canada, the Russian 
Ministry of Science, and the U.S.\@ Department of Energy.



\begin{thebibliography}{49}
\expandafter\ifx\csname natexlab\endcsname\relax\def\natexlab#1{#1}\fi
\expandafter\ifx\csname bibnamefont\endcsname\relax
  \def\bibnamefont#1{#1}\fi
\expandafter\ifx\csname bibfnamefont\endcsname\relax
 \def\bibfnamefont#1{#1}\fi
 \expandafter\ifx\csname citenamefont\endcsname\relax
  \def\citenamefont#1{#1}\fi
\expandafter\ifx\csname url\endcsname\relax
  \def\url#1{\texttt{#1}}\fi
\expandafter\ifx\csname urlprefix\endcsname\relax\def\urlprefix{URL }\fi
\providecommand{\bibinfo}[2]{#2}
\providecommand{\eprint}[2][]{\url{#2}}

\bibitem{Fetscher:1986}
W.~Fetscher, H.-J.~Gerber, and K.F.~Johnson, Phys. Lett. \textbf{B173}, 102 (1986).
\bibitem{PDG}
W.~Fetscher and H.J.~Gerber, in Review of Particle
Physics, K. Nakamura \textit{et al.} (Particle Data Group),
J.~Phys.~G~\textbf{37}, 075021 (2010), p. 521.
\bibitem{Michel50}
L.~Michel, Proc. Phys. Soc. \textbf{A63}, 514 (1950);
C.~Bouchiat and L.~Michel, Phys. Rev. \textbf{106}, 170 (1957);
T.~Kinoshita and A.~Sirlin, Phys. Rev. \textbf{107}, 593 (1957);
T.~Kinoshita and A.~Sirlin, Phys. Rev. \textbf{108}, 844 (1957).
\bibitem{Chizhov94}
M. V.~Chizhov, Mod. Phys. Lett. \textbf{A8}, 2753 (1993).
\bibitem{Chizhov_Review}
	M. V.~Chizhov, Physics of Particles and Nuclei \textbf{42}, 93 (2011).
\bibitem{musser:2005}
J.R.~Musser {\it et al.} (TWIST Collaboration), Phys. Rev. Lett. \textbf{94}, 101805 (2005).
\bibitem{gaponenko:2005}
A.~Gaponenko {\it et al.} (TWIST Collaboration), Phys. Rev. D \textbf{71}, 071101(R) (2005).
\bibitem{Robpaper}
R.P. MacDonald {\it et al.} (TWIST Collaboration), Phys. Rev. D \textbf{78}, 032010 (2008).
\bibitem{mischke:2011}
	R.~Bayes {\it et al.} (TWIST Collaboration), Phys. Rev. Lett. \textbf{106}, 041804 (2011).
\bibitem{Jamespmuxi}
	J.F.~Bueno {\it et al.} (TWIST Collaboration), Phys. Rev. D \textbf{84}, 032005 (2011); \textbf{85}, 039908(E) (2012).
\bibitem{gagliardi:2005}
C.A.~Gagliardi, R.E.~Tribble, and N.J.~Williams, Phys. Rev. D \textbf{72}, 073002 (2005).
\bibitem{TWIST_Apparatus:2005}
	R.S.~Henderson {\it et al.}, Nucl. Instrum. Methods A \textbf{548}, 306 (2005).
\bibitem{Davydov}
	Y.~Davydov {\it et al.}, Nucl. Instrum. Methods A \textbf{461}, 68 (2001).
\bibitem{JamesPRB}
	J.F. Bueno {\it et al.}, Phys. Rev. B \textbf{83}, 205121 (2011).
\bibitem{opera}
	\textsc{Opera} 3D Simulation Software, Vector Fields.
\bibitem{Hu:2006}
J.~Hu {\it et al.}, Nucl. Instrum. Methods A \textbf{566}, 563 (2006).
\bibitem{AndreiPhD}
A.~Gaponenko, arXiv:1104.2914 [hep-ex].
\bibitem{Geant321}
	\textsc{geant} Detector Description and Simulation Tool, CERN Application Software Group, Geneva, Switzerland, 1994, Version 3.21.
\bibitem{Garfield}
	R.~Veenhof, \textsc{Garfield}: Simulation of gaseous detectors, CERN program Library long writeup W5050 (2003) (unpublished).
\bibitem{Arbuzov}
	A.~B.~Arbuzov, Phys. Lett. \textbf{B524}, 99 (2002);
	A.~Arbuzov, A.~Czarnecki, and A.~Gaponenko, Phys. Rev. D \textbf{65}, 113006 (2002);
	A.~Arbuzov and K.~Melnikov, Phys. Rev. D \textbf{66}, 093003 (2002);
	A.~Arbuzov, J. High Energy Phys. \textbf{2003}, 063 (2003).
\bibitem[{\citenamefont{Davydychev et~al.}(2001)\citenamefont{Davydychev,
	Schilcher, and Spiesberger}}]{davydychev01:hadronic_mudecay}
	\bibinfo{author}{\bibfnamefont{A.}~\bibnamefont{Davydychev}},
	\bibinfo{author}{\bibfnamefont{K.}~\bibnamefont{Schilcher}},
	\bibnamefont{and}
	\bibinfo{author}{\bibfnamefont{H.}~\bibnamefont{Spiesberger}},
	\bibinfo{journal}{Eur. Phys. J. C} \textbf{\bibinfo{volume}{19}},
	\bibinfo{pages}{99} (\bibinfo{year}{2001}).
\bibitem{F_James}
	F.~James, Nucl. Instrum. and Methods in Phys. Research \textbf{211}, 145 (1983).
\bibitem{Lutz}
	G.~Lutz, Nucl. Instrum. and Methods A \textbf{273}, 349 (1988).
\bibitem{PDG Thru Matter}
H.~Bichsel, D.E.~Groom, and S.R.~Klein, in Review of Particle
Physics, K. Nakamura \textit{et al.} (Particle Data Group),
J.~Phys.~G~\textbf{37}, 075021 (2010), p. 285.
\bibitem{MINUIT}
	F.~James, \textsc{MINUIT}: Function Minimization and Error Analysis, CERN Program Library Entry D 506 (1994).
\bibitem{AlexNIM}
	A.~Grossheim, J.~Hu, and A.~Olin, Nucl. Instrum. Methods A \textbf{623}, 954 (2010).
\bibitem{Berger}
	M.J.~Berger and S.M.~Seltzer, National Academy of Sciences Report No. 39, 1964.
\bibitem{Anastasiou}
	C. Anastasiou, K. Melnikov, and F. Petriello, J. High Energy Phys. \textbf{09}, 14 (2007).
\bibitem{Barlow}
	R. Barlow and C. Beeston, Computer Physics Communications \textbf{77}, 219 (1993).
\bibitem{fron59}{C. Fronsdal and H. \"Uberall, Phys. Rev. \textbf{113}, 654 (1959).}
\bibitem{burkard:1985}
H.~Burkard {\it et~al.}, Physics Letters \textbf{B160}, 343 (1985).
\bibitem{Jodidio}
A.~Jodidio {\it et al.}, Phys. Rev. D \textbf{34}, 1967 (1986);
\textbf{37}, 237(E) (1988).
\bibitem{SuperK}
Y.~Fukuda {\it et al.}, Phys. Rev. Lett. \textbf{81}, 1562 (1998).
\bibitem{SNO}
Q. R. Ahmad {\it et al.}, Phys. Rev. Lett. \textbf{89}, 011301 (2002).
\bibitem{Doi}
M.~Doi, T.~Kotani, and H.~Nishiura, Prog. Theor. Phys. \textbf{118}, 1069 (2007); \textbf{122}, 805(E) (2009).
\bibitem{Voloshin}
M.~B.~Voloshin, Phys. Lett. B. \textbf{283}, 120 (1992).
\bibitem{Williams}
K.S.~ Williams, private communication (2006).
\bibitem{Derenzo}
S.E.~Derenzo, Phys. Rev. \textbf{181}, 1854 (1969).
\bibitem{Balke}
B.~Balke {\it et al.}, Phys. Rev. D \textbf{37}, 587 (1988).

\end{thebibliography}
\end{document}